\documentclass[a4paper,onecolumn,11pt,accepted=2026-03-25]{quantumarticle}
\pdfoutput=1
\usepackage{bm} 
\usepackage{amsmath} 
\usepackage{amssymb} 
\usepackage{amsthm} 
\usepackage{amsfonts} 
\usepackage[boxed]{algorithm}
\usepackage[noend]{algpseudocode}
\usepackage{physics}
\usepackage{braket}

\newcommand{\labs}{\left|}
\newcommand{\rabs}{\right|}
\newcommand{\lnorm}{\left\|}
\newcommand{\rnorm}{\right\|}
\newcommand{\lcurly}{\left\{}
\newcommand{\rcurly}{\right\}}
\newcommand{\lbrac}{\left[}
\newcommand{\rbrac}{\right]}
\newcommand{\lpar}{\left(}
\newcommand{\rpar}{\right)}

\algnewcommand\algorithmicinput{\textbf{Input:}}
\algnewcommand\Input{\item[\algorithmicinput]}
\algnewcommand\algorithmicoutput{\textbf{Output:}}
\algnewcommand\Output{\item[\algorithmicoutput]}
\algnewcommand\algorithmicalgorithm{\textbf{Algorithm:}}
\algnewcommand\Algorithm{\item[\algorithmicalgorithm]}

\newtheorem{theorem}{Theorem}[section]
\newtheorem{lemma}{Lemma}[section]
\newtheorem{proposition}{Proposition}[section]
\newtheorem{corollary}{Corollary}[section]

\newtheorem{definition}{Definition}[section]
\newtheorem{problem}{Problem}[section]

\DeclareMathOperator{\poly}{poly}
\DeclareMathOperator{\polylog}{polylog}

\DeclareMathOperator{\gap}{\Delta}
\DeclareMathOperator{\sgn}{sgn}

\newcommand\vecd{\boldsymbol{\mathrm{d}}}
\newcommand\vece{\boldsymbol{\mathrm{e}}}
\newcommand\vecf{\boldsymbol{\mathrm{f}}}

\newcommand\vecv{\boldsymbol{\mathrm{v}}}

\newcommand\vecx{\boldsymbol{\mathrm{x}}}
\newcommand\vecy{\boldsymbol{\mathrm{y}}}
\newcommand\vecz{\boldsymbol{\mathrm{z}}}

\newcommand\matA{\boldsymbol{\mathrm{A}}}
\newcommand\matB{\boldsymbol{\mathrm{B}}}

\newcommand\matD{\boldsymbol{\mathrm{D}}}

\newcommand\matH{\boldsymbol{\mathrm{H}}}
\newcommand\matI{\boldsymbol{\mathrm{I}}}

\newcommand\matO{\boldsymbol{\mathrm{O}}}
\newcommand\matP{\boldsymbol{\mathrm{P}}}

\newcommand\matU{\boldsymbol{\mathrm{U}}}

\newcommand\matW{\boldsymbol{\mathrm{W}}}

\newcommand\matZ{\boldsymbol{\mathrm{Z}}}

\newcommand\matPi{\boldsymbol{\mathrm{\Pi}}}

\newcommand\matAtilde{\widetilde{\boldsymbol{\mathrm{A}}}}

\newcommand\matHtilde{\widetilde{\boldsymbol{\mathrm{H}}}}

\newcommand\matHdagH{\boldsymbol{\mathrm{H}^\dagger\mathrm{H}}}
\newcommand\matHdagHtilde{\widetilde{\boldsymbol{\mathrm{H}^\dagger\mathrm{H}}}}

\newcommand{\QSMIN}{\mathsf{QSMIN}}
\newcommand{\QCOUNT}{\mathsf{QCOUNT}}

\newcommand{\QGAPconst}{\mathsf{QGAPconst}}

\newcommand{\QENC}{\mathsf{QENC}}
\newcommand{\QSHIFT}{\mathsf{QSHIFT}}
\newcommand{\QEIG}{\mathsf{QEIG}}
\newcommand{\qcost}{\mathcal{C}}

\usepackage[utf8]{inputenc}
\usepackage[english]{babel}
\usepackage[T1]{fontenc}
\usepackage{listings}
\usepackage{enumerate}
\usepackage[numbers]{natbib}
\usepackage{graphicx}
\usepackage{xcolor}
\usepackage{hyperref} 
\usepackage{cleveref}
\usepackage{setspace}
\hypersetup{
    colorlinks,
}
\usepackage[resetlabels]{multibib}
\newcites{article}{article references}

\begin{document}

\title{Spectral Gaps with Quantum Counting Queries and Oblivious State Preparation}

\author{Almudena Carrera Vazquez}
\affiliation{IBM Research, R\"uschlikon, Switzerland}
\email{acv@zurich.ibm.com}
\orcid{0000-0001-8033-979X}
\author{Aleksandros Sobczyk}
\affiliation{IBM Research, R\"uschlikon, Switzerland}
\affiliation{ETH Zurich, Zurich, Switzerland}
\thanks{the author contributed to this work exclusively while still at IBM Research and ETH Zurich. Current affiliation: Huawei Zurich Research Center.}
\email{sobczykalek@gmail.com}
\orcid{0000-0002-1602-8329}

\maketitle

\begin{abstract}
    Approximating the $k$-th spectral gap $\gap_k=|\lambda_k-\lambda_{k+1}|$ and the corresponding midpoint $\mu_k=\frac{\lambda_k+\lambda_{k+1}}{2}$ of an $N\times N$ Hermitian matrix with eigenvalues $\lambda_1\geq\lambda_2\geq\ldots\geq\lambda_N$,  is an important special case of the eigenproblem with numerous applications in science and engineering. 
    In this work, we present a quantum algorithm which approximates these values up to additive error $\epsilon\gap_k$ using a logarithmic number of qubits. Notably, in the QRAM model, its total complexity (queries and gates) is bounded by $O\lpar \frac{N^2}{\epsilon^{2}\gap_k^2}\polylog\lpar N,\frac{1}{\gap_k},\frac{1}{\epsilon},\frac{1}{\delta}\rpar\rpar$, where $\epsilon,\delta\in(0,1)$ are the accuracy and the failure probability, respectively.  
    For large gaps $\gap_k$, this provides a speed-up against the best-known complexities of classical algorithms, namely, $O \lpar N^{\omega}\polylog \lpar N,\frac{1}{\gap_k},\frac{1}{\epsilon}\rpar\rpar$, where $\omega\lesssim 2.371$ is the matrix multiplication exponent. 
    A key technical step in the analysis is the preparation of a suitable random initial state, which ultimately allows us to efficiently count the number of eigenvalues that are smaller than a threshold, while maintaining a quadratic complexity in $N$.
    In the black-box access model, we also report an $\Omega(N^2)$ query lower bound for deciding the existence of a spectral gap in a binary (albeit non-symmetric) matrix.
\end{abstract}


\newpage
\section{Introduction}
Eigenvalue problems lie in the heart of computational science, engineering, and machine learning. Given an $N\times N$ input matrix $\matA$, the goal is to compute (a subset of) the \emph{eigenvalues} $\lambda_i$ and/or the \emph{eigenvectors} $\vecv_i$, for $i=1,\ldots,N$, which satisfy:
\begin{align*}
    \matA\vecv_i=\lambda_i\vecv_i.
\end{align*}

In this work we focus on algorithms for an important special case of the eigenproblem, namely, the approximation of the so-called \emph{spectral gap} (or \emph{eigengap}) and the corresponding midpoint between pairs of consecutive eigenvalues of \emph{Hermitian} matrices, which we formally define as follows.
\begin{problem}[Spectral gap and midpoint approximation]
\label{problem:gap}
Let $\matH$ be an $N\times N$ Hermitian matrix, with eigenvalues $ \lambda_1\geq\lambda_2\geq\ldots\geq\lambda_N $, and $\gap_k:=|\lambda_k-\lambda_{k+1}|$ be the $k$-th spectral gap and $\mu_k:=\frac{\lambda_k+\lambda_{k+1}}{2}$ is the corresponding midpoint between $\lambda_k$ and $\lambda_{k+1}$. Given $\matH$, an integer $k\in[N-1]$, and an accuracy $\epsilon\in(0,1)$, compute two values $\widetilde \gap_k, \widetilde \mu_k$ such that: 
\begin{align*}
    \widetilde \gap_k\in (1\pm\epsilon)\gap_k, \text{\quad and }\quad \widetilde \mu_k\in \mu_k\pm \epsilon\tfrac{\gap_k}{2}\subseteq[\lambda_{k+1},\lambda_{k}].
\end{align*}
No prior information regarding the eigenvalues or the eigenvectors of $\matH$ is assumed.     
\end{problem}
Problem \ref{problem:gap} has numerous applications in different areas. In numerical analysis, estimating spectral gaps is a key subroutine of so-called \emph{spectral divide-and-conquer} algorithms for diagonalizing matrices \cite{beavers1974new,nakatsukasa2013stable,banks2023pseudospectral,shah2025hermitian}. These methods rely on finding a ``splitting'' point $\mu$ that does not coincide with an eigenvalue; our midpoint estimator $\mu_k = (\lambda_k + \lambda_{k+1})/2$ provides a natural and
well-conditioned choice for such a split, as it maximizes the distance to the nearest eigenvalues.
In data analytics, a large gap in the spectrum of a data matrix indicates a separation between the actual information and the ``noise'' in the data. 
This phenomenon is often used in Principal Component Analysis \cite{jolliffe2002principal,scholkopf1997kernel} to identify good splitting points, or ``elbows'', for dimensionality reduction, data visualization, and noise filtering. By directly targeting the $k$-th gap, our algorithm enables the automated detection of these elbows without requiring the explicit computation of (potentially large) subsets of singular values or a full singular value decomposition (SVD).
In spectral graph theory, gaps between the eigenvalues of the graph Laplacian reveal connectivity properties, and they can be used, for example, to identify communities, \cite{von2007tutorial,cheeger1970lower,alon1985lambda1,louis2012many}. 

In the study of quantum phase transitions, the closing of the $k$-th gap, where $\lambda_{k+1} - \lambda_k \to 0$ as the system size increases, serves as a key indicator of topological or magnetic phase transitions \cite{sachdev2011quantum}. 
In the context of quantum computing, the size of the spectral gap also determines the runtime of adiabatic state preparation, since the adiabatic theorem requires the evolution time to scale inversely with (a power of) the minimum spectral gap encountered \cite{farhi2001quantum}. Last but not least, spectral gaps are also important in electronic structure calculations, for example, to estimate the energy gap (Fermi level) that separates the occupied and unoccupied states of the underlying system \cite{zhou2006self,kresse1996efficient,vandevondele2012linear,lin2019numerical}. In that direction, our gap estimator could be combined with the techniques of  \cite{ko2023implementation} to calculate the electron density of a system at zero temperature.

The aim of this work is to provide a detailed analysis of quantum algorithms for solving Problem \ref{problem:gap}, aiming for provable approximation guarantees and end-to-end \emph{worst-case} complexity bounds. 

\subsection{Existing algorithms and complexity bounds}

Here we give a brief overview of the related algorithms literature. We first recall that, from the Abel-Ruffini theorem, computing eigenvalues of general matrices is intractable in the algebraic model; see e.g.  \cite{ziegler2004computability}. Besides trivial cases, such as triangular matrices, algorithms for eigenvalue problems in general return approximate solutions. The computational complexity depends on several different factors, namely, the underlying model of computation, the properties of the input matrix (for example, Hermitian, sparse, structured, or low-rank), the quantities that need to be computed (subset of eigenvalues/eigenvectors), the accuracy, and, for randomized algorithms, the success probability.

In the classical (non-quantum) regime,
recent works have established that both the arithmetic and boolean worst-case complexity of \emph{diagonalization} (approximation of \emph{all} the eigenvalues and eigenvectors), is upper bounded by $\widetilde O(N^{\omega})$ operations \cite{banks2023pseudospectral,sobczyk2024invariant,sobczyk2024deterministic,demmel2024generalized,demmel2026fast,shah2025hermitian}.
Here $\omega$ is the matrix multiplication exponent, currently upper bounded by $\omega<2.371339$ \cite{alman2025more}, and $\widetilde O$ hides polylogarithmic factors in $N$ and potentially other parameters, such as the accuracy. In terms of solving Problem \ref{problem:gap} \emph{per-se}, \cite{sobczyk2024deterministic}   reports a complexity of $O \lpar N^{\omega}\polylog \lpar N,\frac{1}{\gap_k},\frac{1}{\epsilon}\rpar\rpar$ operations in the Boolean RAM model.

Beyond the matrix multiplication-type worst-case complexity, fast algorithms have been studied for special classes of matrices. For instance, the eigenvalues of banded symmetric matrices with $k$ bands can be computed in $\widetilde O(N^2k^{\omega-2})$ time via tridiagonal reduction \cite{sobczyk2024deterministic,gu1995divide}. For symmetric diagonally dominant matrices (SDD) and graph Laplacians, the smallest $k$ eigenvalues of an $N\times N$ SDD matrix with $M$ non-zeros can be approximated in $\widetilde O(M+kN/\epsilon^2)$ time \cite{spielman2014nearly,koutis2015faster}. This is faster than the aforementioned $\widetilde O(N^\omega)$ bounds even for $k\in \Theta(N)$ and $M\in\Theta(N^2)$, albeit with increased dependence on the accuracy $\epsilon$ (polynomial instead of logarithmic).  The recent work of \cite{benzi2024estimation} described an elegant approach specifically targeting spectral gaps of general (non-SDD) sparse symmetric matrices based on stochastic trace estimation \cite{hutchinson1989stochastic,avron2011randomized,roosta2015improved,cortinovis2022randomized}, albeit, the current worst-case complexity analysis can grow larger than $\widetilde O(N^\omega)$.

The literature on quantum algorithms for eigenvalue problems is vast, and, to date, it keeps growing rapidly. The majority of existing works has focused on the following  problems: (i) estimation of extremal eigenvalues, i.e., the ground state energy, (ii) approximating an eigenvalue, given access to the corresponding eigenvector, or (iii) preparation of quantum states that encode eigenvalues as amplitudes. Such methods include the so-called \emph{Variational Quantum Eigensolvers} \cite{tilly2022variational,peruzzo2014variational,mcclean2016theory}, and methods based on \emph{Quantum Phase Estimation} (QPE)  \cite{kitaev1995,abrams1999quantum,dorner2009optimal}, such as \cite{Lin2020nearoptimalground,low2024quantum,kerzner2023square,kerenidis2016quantum,shao2022computing,parker2020quantum,ding2024quantum}.

Here we highlight recent works that can be considered for solving Problem 
\ref{problem:gap}, with explicit worst-case complexity bounds. Low and Su \cite{low2024quantum} (see e.g. Theorem 3) recently proved  sharp quantum query bounds for estimating any eigenvalue of general matrices. The query bound is, $O((a\epsilon)^{-1}\log(1/\delta))$ for Hermitian matrices, where $a$ is the block-encoding scaling factor. The algorithm assumes as input an initial state that is $\epsilon$-close (in the Euclidean distance) to the corresponding eigenvector. Such a state is hard to obtain in the  general case, as it entails to approximate an (unknown) eigenvector.

State preparation assumptions can be relaxed for the problem of estimating the smallest eigenvalues. Recent works of  \cite{Lin2020nearoptimalground,ding2023simultaneous,ding2024quantum,kerzner2023square} suggest that $1/\poly(N)$ overlap between the initial state and the target bottom-$k$ states is sufficient. If an initial state that satisfies this requirement is readily available, then a straightforward approach to solve Problem \ref{problem:gap} is to directly \cite{ding2023simultaneous,ding2024quantum} to approximate all $N-k$ smallest eigenvalues, in a ``brute-force'' manner. 
As it is clearly noted in these works, however, this approach is targeted for approximating few eigenvalues close to the end of the spectrum, in this case, when $N-k$ is small. Moreover, beyond the initial state preparation assumptions, it is required that the $N-k$ smallest eigenvalues are strictly separated by a well-defined minimum gap. For large $N-k\in\Theta(N)$, and for general Hermitian matrices, it becomes challenging to ensure or to override these assumptions and to guarantee competitive end-to-end complexity bounds in the worst-case.

End-to-end worst-case complexity bounds were proved in recent works by Apers and de Wolf \cite{apers2022quantum}, and by Chen, Gily\'en and de Wolf \cite{chen2025quantum}. The former  showed how to approximate the top-$k$ eigenvalues and eigenvectors of SDD matrices with $M$ non-zeros in  $\widetilde O(\sqrt{MN}+kN/\epsilon^2)$ time in the QRAM model. It is faster than the $O(M+kN/\epsilon^2)$ classical SDD algorithms for small $k$ and large $M$. For \emph{general} square matrices, \cite{chen2025quantum} showed how to approximate the top-$k$ eigenvectors in  $\widetilde O(kN^{1.5+o(1)})$, using quantum versions of the power method. As also noted by the authors, however, the algorithm is  less competitive for large $k\in\Theta(N)$. 

As a last remark, we note that obtaining initial states which sufficiently approximate the target eigenvectors is an interesting research topic on its own. For certain applications, an initial state can be obtained by solving an ``easier'' or ``smaller'' version of the problem as a pre-processing step. 
As \cite{Lin2020nearoptimalground} notes, this is particularly interesting for quantum chemistry applications \cite{kivlichan2018quantum}. Random states are often recommended for the general case, but they can be also expensive to prepare, as we also discuss here.

\subsection{Contributions and methodology}

Our main contribution is an algorithm (Algorithm \ref{algorithm:qeig}) to solve Problem \ref{problem:gap}. In a high-level, it consists of two phases:
\begin{itemize}
    \item In the first phase (Alg.\@ \ref{algorithm:qgap_const}), the spectral gap $\gap_k$ and the midpoint $\mu_k$ are computed to \emph{constant accuracy} (i.e., a crude initial approximation).
    \item The second phase (Alg.\@ \ref{algorithm:qeig}) uses the values returned in the first phase in order to perform a refined search in the restricted domain and increase the solution accuracy.
\end{itemize} 
Both phases use the following steps to locate gaps and/or eigenvalues:
\begin{enumerate}
    \item Encode the given Hermitian matrix in a quantum circuit.
    \item Split the spectrum in two parts, by shifting the diagonal elements to a carefully chosen point.
    \item Compute the spectral projector on each part of the spectrum with a polynomial approximation of the sign function.
    \item Compute the trace of the projector.
    \item Repeat 2-4 as a search mechanism (``counting queries'') over a grid to locate eigenvalues and/or gaps.
\end{enumerate}
For Step 1, we use the block-encoding framework \cite{gilyen2019quantum,chakraborty2019power}, which has recently received significant attention for matrix problems. We use the following definition.
\begin{definition}[Block-Encoding]
    \label{definition:block_encoding}
    Let $\matA\in\mathbb{C}^{N\times N}$ be an $s$-qubit matrix, $a\geq\|\matA\|, \epsilon>0$, and $b\in\mathbb{N}$.
    An $(s+b)$-Qubit Unitary $\matU_{\matA}$ is an $(a,b,\epsilon)$-Block-Encoding of $\matA$, which is also abbreviated as $(a,b,\epsilon)$-QUBE for short, if
    \begin{equation}
        \lnorm 
        \matA- a\matAtilde
        \rnorm
        \leq 
        \epsilon,
    \end{equation}
    where $\matAtilde:=(\bra{0}^{\otimes b} 
        \otimes \matI)\matU_{\matA}(\ket{0}^{\otimes b} \otimes \matI)$.
\end{definition}

Our main result is summarized in the following theorem.
\begin{theorem}[Informal, see Theorem \ref{theorem:alg_qeig}]
    \label{theorem:alg_qeig_informal}
    Let $\matH$ be an $N\times N$ Hermitian matrix  with eigenvalues $ -1/2\leq\lambda_N\leq\lambda_{N-1}\leq\ldots\leq\lambda_1 \leq 1/2$, $\gap_k:=|\lambda_k-\lambda_{k+1}|$, and $\mu_k:=\frac{\lambda_k+\lambda_{k+1}}{2}$, and let $\qcost(\matU_{\matH};\epsilon_{\rm enc})$ denote the cost (gate complexity) of the unitary $\matU_{\matH}$ which forms an $(a,b,\epsilon_{\rm enc})$-block-encoding of $\matH$.
    The proposed Algorithm \ref{algorithm:qeig} takes as input $\matH$, an integer $k\in[N-1]$, an accuracy $\epsilon\in(0,1)$, and a failure probability parameter $\delta\in(0,1/2)$, and it returns  values $\widetilde\lambda_k,  \widetilde\lambda_{k+1},  \widetilde\mu_k,$ and $\widetilde\gap_k$, which satisfy:
    \begin{align*}        \widetilde\lambda_{k+1}&\in[\lambda_{k+1},\lambda_{k+1}+\epsilon\tfrac{\gap_k}{2}],
        &
    \widetilde\lambda_k&\in[\lambda_k-\epsilon\tfrac{\gap_k}{2},\lambda_k],
        &
    \widetilde\mu_k &\in \mu_k\pm \epsilon\tfrac{\gap_k}{2},
        &
    \widetilde\gap_k&\in(1\pm\epsilon)\gap_k.
    \end{align*}
    The total complexity, including all queries to unitaries and auxiliary gates, is bounded by:
        \begin{align}
        O\left(
            \left[
            N^2
            +
            N\cdot \tfrac{a^2 b\qcost(\matU_{\matH};\epsilon_{\rm enc})}{\epsilon^2\gap_k^2}
            \right]
            \polylog\left(N,a,\epsilon^{-1},\gap_k^{-1},\delta^{-1}\right)
        \right),
        \end{align}
        where $\epsilon_{\rm enc}\in \Theta\left(
        \frac{\epsilon^2\gap_k^2}{a^2N}
        \right)$.
\end{theorem}
Notably, in the QRAM model, the complexity becomes  $\widetilde O(N^2\gap_k^{-2}\epsilon^{-2})$. This is faster than the $\widetilde O(N^\omega)$ arithmetic complexity of the best-known classical algorithms, for all matrices with sufficiently large gaps. This provides evidence that quantum algorithms can beat ``matrix multiplication-time'' for these problems, at least for the current bound of $\omega$.

While the general idea is simple, there are several non-trivial parts that need to be addressed to obtain competitive end-to-end complexities. Two major challenges, which can be of independent interest, were the choice of the splitting points, and the preparation of appropriate initial states, which we discuss further below.

\subsubsection*{Informed choice of splitting points via condition number estimation} As mentioned, the splitting points to split the spectrum and count eigenvalues. If the chosen point is close to an actual eigenvalue, the condition number of the shifted matrix can grow arbitrarily, and it becomes prohibitively expensive to approximate the corresponding spectral projector. 
Indeed, the convergence of iterative methods, such as the Newton-Schultz iteration or other rational approximations \cite{kenney1991rational}, scales as $\Omega(1/\gap)$, where $\gap$ is the distance of the splitting point from the closest eigenvalue \cite{gilyen2019quantum}. For this reason, many existing algorithms that follow this approach require prior knowledge of $\gap$ (or at least a suitable lower bound), which is generally not trivial to obtain. 

To avoid this problem, we use a dedicated subroutine which tests whether the gap $\gap$ is larger than a well-defined threshold, before executing any counting query. 
This is achieved by conveniently applying the ground-energy estimation algorithm of \cite{Lin2020nearoptimalground} in order to estimate the smallest singular value of the matrix, after shifting its diagonal elements to the chosen splitting point. 
If this singular value is ``large enough'', it means that we can safely count eigenvalues. If not, we keep searching for smaller gaps. This way we have full control of the value of $\gap$, and, consequently, of the cost of eigenvalue counting.

\subsubsection*{Initial state preparation}
In order to execute our singular value estimation subroutine efficiently, we need to address a second major challenge, which involves the preparation of an appropriate initial state. The algorithm of \cite{Lin2020nearoptimalground} requires as input an initial state $\ket{x}$ which will overlap appropriately with the corresponding singular vector. 
In particular, this target singular vector can be any fixed state $\ket{y}$, and we need to ensure that the initial guess $\ket{x}$
will satisfy \begin{align}
    \abs{\braket{x|y}}\geq \frac{1}{N^{O(1)}}. \label{eq:initial_state_requirements}
\end{align}
As the authors of \cite{Lin2020nearoptimalground,ding2023simultaneous} note, such an initial state can be efficiently prepared for certain classes of problems. For instance,
if $\ket{y}$ is a binary string with exactly one nonzero entry, a Hadamard state is sufficient, or, in quantum chemistry applications, the Hartree-Fock state can be pre-computed as a ``cheap'' initial approximation to the ground state \cite{kivlichan2018quantum}.

In the general case, where $\ket{y}$ can be any arbitrary state without any explicit prior information, the analysis becomes more involved. In the context of QPE, a common recommendation is to use a random state. 
However, if the random distribution is not chosen carefully, it can be harmful rather than beneficial. Contributing to this direction,
in Section \ref{section:state_preparation} we analyze a randomized state preparation method, which achieves all the desired properties. It is \emph{oblivious} to the target state, in a sense that it guarantees the desired overlap $1/N$, for any fixed target state, with probability at least $3/5$. Our construction is an ensemble of a random Rademacher state, together with a randomized Hadamard transform \cite{ailon2009fast,nguyen2009fast,tropp2011improved}. The latter is a well-established tool in Randomized Numerical Linear Algebra, and it is crucial for our analysis do to its ``flattening'' properties, commonly attributed as a consequence of the Heisenberg principle. 

The size of the final quantum circuit scales as $O(N)$, and it requires a total of $O(N)$ random classical bits to prepare. While this cost is prohibitive for applications that target exponential quantum speed-ups, it is ``sufficiently cheap'' for our analysis, and it can be useful for other works that target polynomial quantum speed-ups.

\subsubsection*{Lower bounds}
The aforementioned upper bounds are also supported with a $\Omega(N^2)$ query lower bound for the problem of deciding whether there exists a gap between two consecutive eigenvalues of a (non-symmetric) binary matrix in the black-box access model. The techniques used here are based on  ``majority'' arguments, inspired by \cite{ambainis2000quantum,dorn2009quantum,gribling2024optimal}. 
Evidently, these lower bounds are not tight, as they do not match the reported upper bound. In order to close this complexity gap, we think that one should use entirely different techniques, both for the upper and the lower bounds. This is left as a major open question.

\subsection{Notation}
$\mathcal{H}_S$ denotes an explicit Hilbert space and $\mathcal{H}_A$ an ancillary space for post-selection. When we want to emphasize its dimension $d$, we write $\mathcal{H}_A\cong \mathbb{C}^d$. The total Hilbert space is given by $\mathcal{H}=\mathcal{H}_S\otimes\mathcal{H}_A$. Matrices and vectors are denoted with bold capital and small letters, respectively. 
For a matrix $\matA$, $\matA^\dagger$ is the conjugate transpose, $\lVert \matA\rVert$ denotes its spectral norm, $\lVert \matA\rVert_F$ its Frobenius norm, and $\Lambda(\matA)$ denotes its spectrum. $\qcost(\matU)$ denotes the gate complexity of the unitary $\matU$, and we will write $\qcost(\matU;\epsilon,\delta)$ when the complexity of $\matU$ is parametrized by $\epsilon,\delta$.

\subsection{Model of computation}
The complexity of a quantum algorithm is two-fold. We define the \emph{query complexity}  to be the total number of calls to the block-encoding of the original matrix. The \emph{gate complexity} is the total number of elementary one- and two-qubit quantum gates, i.e., the size of the quantum circuit.  
For all algorithms, we explicitly mention both the query and the gate complexities, taking into account all the involved parameters, such as the matrix size, the accuracy, and the success probabilities.
For our state-preparation analysis, we also assume access to a (classical) random number generator that produces independent, uniformly random bits.  

In Section \ref{section:qram}, we specialize the analysis in the so-called Quantum-RAM (or QRAM) model \cite{giovannetti2008quantum}. In this model the quantum processor has access to a memory that stores classical bits. 
These bits can be read in superposition, but we can only write zeros and ones, just as in the classical case. 
This model is not necessary for our main analysis, but we dedicate Section \ref{section:qram} to it since it is used in theoretical works, e.g., for graph and matrix computations \cite{apers2022quantum,chen2025quantum,kerenidis2016quantum}. For a discussion on QRAM architectures we refer to \cite{giovannetti2008architectures,jaques2025qram}.

\subsection{Outline}
\Cref{section:main_subroutines} contains the analysis of the main subroutines that are used by the main algorithms, namely, for state preparation, computing the minimum singular value, and counting eigenvalues that are smaller than a threshold. \Cref{section:main_algorithm} contains the main algorithm for computing spectral gaps and eigenvalue pairs. In \Cref{section:qram} we report the analysis in the QRAM model. In \Cref{section:lower_bounds} we prove  a query lower bound for deciding the existence of a spectral gap in binary matrices in the black-box access model. We finally provide some concluding remarks in \Cref{section:conclusion}.

\acknowledgements{We are grateful to Ivano Tavernelli and Simon Apers for valuable discussions.}

\section{Main Subroutines}
\label{section:main_subroutines}
Here we describe the main subroutines that are used as building blocks for our main algorithms. Some of the techniques reported here rely on existing results on block-encodings and quantum eigenvalue transformation from \cite{gilyen2019quantum,rall2020,Lin2020nearoptimalground,chakraborty2019power}, which are summarized in Appendix \ref{appendix:preliminaries}.

\subsection{Oblivious State Preparation}
\label{section:state_preparation}
In many applications of QPE, it is required to prepare an initial state $\ket{x}$ that is not orthogonal to the target state $\ket{y}$ (e.g. the ground state). A sufficient condition is typically  $\abs{\braket{x|y}}^2\geq \Omega(1/N)$. 
The state $\ket{x}=\ket{+}:=\matH\ket{0}$ is sufficient for many applications, e.g.,  when the desired state is a binary string. For arbitrary $\ket{x}$, however, there is no theoretical guarantee that $\ket{+}$ is not orthogonal to $\ket{x}$. For example, it is known that $\ket{+}$ spans the kernel of a connected graph Laplacian, and it is therefore orthogonal to \emph{every}  eigenvector that corresponds to a nonzero eigenvalue. 

A reasonable approach to circumvent this limitation is to use random states. However, as already mentioned previously, this needs to be handled with care.  
Here we describe a random state preparation, inspired by well-known techniques from randomized  matrix algorithms \cite{ailon2009fast}. Before stating the main result, we define a state-preparation unitary.
\begin{definition}
    \label{definition:state_preparation_unitary}
    A distribution $\matPi$ of random quantum circuits forms a $(\gamma,\delta)$-Oblivious State Preparation, or $(\gamma,\delta)$-OSP, if the unitary $\matU_{S}\sim \matPi$  operates on $O(\log(N))$ qubits and it prepares a state $\ket{x}=\matU_{S}\ket{0}$ such that, for any fixed state $\ket{y}\in\mathbb{C}^N$:
    \begin{align*}
        \Pr\lbrac
            \abs{\braket{x|y}} < \gamma
        \rbrac
        \leq \delta.
    \end{align*}
\end{definition}

\begin{lemma}
    \label{lemma:oblivious_state_preparation}
    Let $\ket{y}\in\mathbb{C}^{N}$ be any fixed state (e.g. the target ground-state). Let $\ket{x}=\matD\matH\ket{z}$. Here $\matD$ is  diagonal with independent $\{\pm 1\}$ diagonal elements, $\matH :=\begin{pmatrix}
        \tfrac{1}{\sqrt{2}} & \tfrac{1}{\sqrt{2}} \\
        \tfrac{1}{\sqrt{2}} & -\tfrac{1}{\sqrt{2}}
    \end{pmatrix}^{\otimes \log(N)}$ is a Hadamard unitary, and $\ket{z}\in\lcurly \pm \frac{1}{\sqrt{N}}\rcurly^{N}$ has independent Rademacher elements. 
    Let $\matZ$ be a unitary such that $\matZ\ket{0}=\ket{z}$. Then, for all $N\geq C$, where $C\approx 22$ is a fixed constant, the distribution of unitaries $\matU_{S}=\matD\matH\matZ$ is an $(\frac{1}{2N},\frac{2}{5})$-OSP, that is, 
    \begin{align*}
        \Pr\left[\labs \braket{x| y}\rabs > \frac{1}{2N} \right] \geq \frac{3}{5},
    \end{align*}
   The state $\ket{x}$ can be prepared with $O(N)$ elementary gates, i.e., $\qcost(\matU_{S})\in O(N)$.
    \begin{proof}
        Let us denote $\vecx=\ket{x}, 
        \vecz=\ket{z}$, and $\vecy=\ket{y}$ to simplify the notation.
        The operator $\matH\matD$ is a well-studied tool in randomized linear algebra (see e.g. \cite{ailon2009fast}), which has the property that it ``flattens'' the elements of any fixed vector with high probability. Indeed, using Hoeffding's inequality, we can see that for any index $i\in[N]$ it holds that
        \begin{align*}
            \Pr[|(\matH\matD\vecy)_i|\geq t] \leq 2\exp(-Nt^2/2).
        \end{align*}
        Taking a union bound for all $i\in[N]$ and setting $t=\sqrt{\frac{2\ln(2N/\delta)}{N}}$, we can conclude that \begin{align*}
            \Pr[\|\matH\matD\vecy\|_{\infty}\geq t] \leq \delta.
        \end{align*}
        Denote $\vecv:=\matH\matD\vecy$. It remains to get a lower bound on the overlap $|\vecx^\top \vecy|=|\vecz^\top\vecv|$. Since $\matH\matD$ is unitary, $\vecv$ is a unit vector with flattened entries, which is the key observation to obtain the desired bound. Let $\vecv_m$ and $\vecv_{m'}$ be the largest and second largest elements of $\vecv$ in modulus, where $m,m'\in[N]$. It holds that $|\vecv_m|^2\in[1/N,t^2]$. For $\vecv_{m'}$, observe that \begin{align*}
            |\vecv_{m'}|^2 &\geq 
            \frac{\|\vecv\|_2^2 - |\vecv_m|^2}{N-1}
            \geq
            \frac{1 - t^2}{N-1}.
        \end{align*}
        For $\delta=1/5$ and $N\gtrsim 22$, we have that $|\vecv_{m'}|^2\geq 1/(2N-2)>1/(2N)$.
        We now condition on the success of the flattening, and proceed to get a bound for the overlap. Let $\vecz_A,\vecz_B,\vecz_C,\vecz_D$ be four vectors that have exactly the same signs at all the positions except $\{m,m'\}$, where in these two positions they have the signs $\{+1,+1\}, \{+1,-1\}, \{-1,+1\},\{-1,-1\}$, respectively. The following relations hold:
        \begingroup
        \allowdisplaybreaks
        \begin{align*}
            |\vecz_A^\top \vecv - \vecz_B^\top\vecv| &\geq \frac{2|\vecv_{m'}|}{\sqrt{N}} 
            &
                &\implies  
                \max\left\{|\vecz_A^\top\vecv|,|\vecz_B^\top\vecv| \right\} \geq \frac{|\vecv_{m'}|}{\sqrt{N}}
                > \frac{1}{2N}
            \\
            |\vecz_A^\top \vecv - \vecz_C^\top\vecv| &\geq \frac{2|\vecv_{m}|}{\sqrt{N}}
            &
                &\implies  
                \max\left\{|\vecz_A^\top\vecv|,|\vecz_C^\top\vecv| \right\} \geq \frac{|\vecv_{m}|}{\sqrt{N}}
                > \frac{1}{2N}
            \\
            |\vecz_A^\top \vecv - \vecz_D^\top\vecv| &\geq \frac{2(|\vecv_{m}|+|\vecv_{m'}|)}{\sqrt{N}}
            &
                &\implies  
                \max\left\{|\vecz_A^\top\vecv|,|\vecz_D^\top\vecv| \right\} \geq \frac{|\vecv_{m}|+|\vecv_{m'}|}{\sqrt{N}}
                > \frac{1}{2N}
            \\
            |\vecz_B^\top \vecv - \vecz_C^\top\vecv| &\geq \frac{2(|\vecv_{m}|+|\vecv_{m'}|)}{\sqrt{N}}
            &
                &\implies  
                \max\left\{|\vecz_B^\top\vecv|,|\vecz_C^\top\vecv| \right\} \geq \frac{|\vecv_{m}|+|\vecv_{m'}|}{\sqrt{N}}
                > \frac{1}{2N}
            \\
            |\vecz_B^\top \vecv - \vecz_D^\top\vecv| &\geq \frac{2|\vecv_{m}|}{\sqrt{N}}
            &
                &\implies  
                \max\left\{|\vecz_B^\top\vecv|,|\vecz_D^\top\vecv| \right\} \geq \frac{|\vecv_{m}|}{\sqrt{N}}
                > \frac{1}{2N}
            \\
            |\vecz_C^\top \vecv - \vecz_D^\top\vecv| &\geq \frac{2|\vecv_{m'}|}{\sqrt{N}}
            &
                &\implies  
                \max\left\{|\vecz_C^\top\vecv|,|\vecz_D^\top\vecv| \right\} \geq \frac{|\vecv_{m'}|}{\sqrt{N}}
                > \frac{1}{2N}.
        \end{align*}
        \endgroup
        
        Now, denote $L:=1/(2N)$, and let $A$ be the event that $|\vecz_A^\top\vecv|> L$, $B:|\vecz_B^\top\vecv|> L$, $C:|\vecz_C^\top\vecv|> L$, and $D:|\vecz_D^\top\vecv|> L$. From the relations above, we can see that (deterministically), we are in one of the following five cases:
        \begingroup
        \allowdisplaybreaks
        \begin{align*}
            &A \wedge B \wedge C \wedge D\\
            \neg &A \implies B\wedge C\wedge D\\
            \neg &B \implies A\wedge C\wedge D\\
            \neg &C \implies A\wedge B\wedge D\\
            \neg &D \implies A\wedge B\wedge C.
        \end{align*}
        \endgroup
        In all cases, at least three out of four vectors $\{\vecz_A,\vecz_B,\vecz_C,\vecz_D\}$ satisfy the desired lower bound (deterministically). Then, since we pick any of the four vectors uniformly at random, the success probability is at least $3/4$. 
        
        Since the flattening succeeds with probability $4/5$, both events succeed with probability at least $\frac{3}{4}\cdot\frac{4}{5}=\frac{3}{5}$, which concludes the proof.
    \end{proof}
\end{lemma}

\subsection{Computing the smallest singular value }\label{section:qsmin}
Here we describe the subroutine which computes the smallest singular value of a block-encoded matrix.
\cite{Lin2020nearoptimalground} gave an algorithm with near optimal dependence on the precision $\epsilon$ to find the smallest eigenvalue of a Hermitian matrix $\matH$.

\begin{theorem}[Ground energy, Imported Thm.~8 from~\cite{Lin2020nearoptimalground}]
    \label{theorem:approximate_ground_energy}
    Let $\matH=\sum_k \lambda_k \ket{\psi_k}\bra{\psi_k}\in\mathbb{C}^{N\times N}$, where $\lambda_k \geq \lambda_{k+1}$, be a Hermitian matrix given through its $(\alpha, m, 0)$-block-encoding $\matU_{\matH}$.
    Let $\ket{\phi_0}$ be an initial state prepared by circuit $\matU_{S}$, with the promise $\abs{\braket{\phi_0|\psi_0}}\geq\gamma$. 
    Then the ground energy $\lambda_{N}$ can be estimated to additive precision $\epsilon$ with probability $1-\delta$ with the following costs:
    \begin{enumerate}
        \item Queries to $\matU_{\matH}$: $O\left(\frac{\alpha}{\gamma\epsilon}\log(\frac{\alpha}{\epsilon})\log(\frac{1}{\gamma})\log(\frac{\log(\alpha/\epsilon)}{\delta})\right)$,
        \item Queries to $\matU_{S}$: $O\left(\frac{1}{\gamma}\log(\frac{\alpha}{\epsilon})\log(\frac{\log(\alpha/\epsilon)}{\delta})\right)$,
        \item Number of qubits: $O\left(\log(N)+m+\log(\frac{1}{\gamma})\right)$,
        \item Other elementary gates: $O\left(\frac{m\alpha}{\gamma\epsilon}\log(\frac{\alpha}{\epsilon})\log(\frac{1}{\gamma})\log(\frac{\log(\alpha/\epsilon)}{\delta})\right)$.
    \end{enumerate}
\end{theorem}
This can be applied appropriately on the Gramian $\matH^\dagger\matH$ to compute the minimum singular value.
\begin{lemma}[$\QSMIN$]
    \label{lemma:qsmin}
    Let $\matH\in\mathbb{C}^{N\times N}$ be a Hermitian matrix, with singular values $0\leq \sigma_N \leq \sigma_{N-1}\leq \ldots \leq \sigma_1\leq 1$. Let $\epsilon_\sigma\in(0,1/2]$ be an accuracy parameter, and assume that we have access to:
    \begin{itemize}
        \item $\matU_{\matH}$, an $(a,b,\epsilon_{\rm enc})$-QUBE for $\matH$,  $\epsilon_{\rm enc}\in[0,\epsilon_{\sigma}/4a]$, with cost $\qcost(\matU_{\matH};\epsilon_{\rm enc})$
        \item $\matPi$, a $(\gamma,\delta_{\rm osp})$-OSP, where $\delta_{\rm osp}\in[0,1/2)$, and $\matU_{S}\sim \matPi$ has cost $\qcost(\matU_{S})$.
    \end{itemize}
    We can compute a boolean string $X\gets \QSMIN(\matU_{\matH},\epsilon_{\sigma},\delta_{\sigma})$, such that $\widetilde\sigma^2:=0.X\in[0,1]$  satisfies $|\widetilde{\sigma}^2-\sigma_N^2|\leq \epsilon_{\sigma}$ with probability at least $1-\delta_{\sigma}$, with the following costs:
    \begin{enumerate}
            \item Queries to $\matU_{\matH}$: $O\left(
                \frac{a^2}{\gamma\epsilon_{\sigma}}
                \log(\frac{1}{\gamma})
                \log(\frac{a}{\epsilon_{\sigma}})
                \log(
                    \frac{
                    \log(a/\epsilon_{\sigma})
                    }
                    {1-2\delta_{\rm osp}}
                )
                \frac{\log(1/\delta_{\sigma})}{1-2\delta_{\rm osp}}
            \right)$,
            \item Queries to random circuits $\matU_{S}\sim \matPi$: $O\left(
                \frac{1}{\gamma}
                \log(\frac{a}{\epsilon_{\sigma}})
                \log(
                    \frac{
                    \log(a/\epsilon_{\sigma})
                    }
                    {1-2\delta_{\rm osp}}
                )
                \frac{\log(1/\delta_{\sigma})}{1-2\delta_{\rm osp}}
            \right)$,
            \item Number of qubits: $O\left(\log(N) +  b + \log(\frac{1}{\gamma})\right)$,
            \item Other elementary gates: $O\left(
                \frac{a^2b}{\gamma\epsilon_{\sigma}}
                \log(\frac{1}{\gamma})
                \log(\frac{a}{\epsilon_{\sigma}})
                \log(
                    \frac{
                    \log(a/\epsilon_{\sigma})
                    }
                    {1-2\delta_{\rm osp}}
                )
                \frac{\log(1/\delta_{\sigma})}{1-2\delta_{\rm osp}}
            \right)$.
        \end{enumerate}
    \begin{proof}
        Given $\matU_{\matH}$, an $(a,b, \epsilon_{\rm enc})$-QUBE for $\matH$ with cost $\qcost(\matU_{\matH};\epsilon_{\rm enc})$, it holds that $\matU_{\matH}^\dagger$ is an $(a,b, \epsilon_{\rm enc})$-QUBE for $\matH^\dagger$, also with cost $\qcost(\matU_{\matH};\epsilon_{\rm enc})$.
        We can combine these two block-encodings as in \cite[Lemma 4]{chakraborty2019power} to create an $(a^2,2b,2a\epsilon_{\rm enc})$-QUBE for $\matHdagH$, denoted by $\matU_{\matHdagH}$,with cost equal to $\qcost(\matU_{\matHdagH};\epsilon_{\rm enc})=2\qcost(\matU_{\matH};\epsilon_{\rm enc})$.
        The minimum eigenvalue of $\matHdagH$ is the square of the minimum singular value of $\matH$. 
        
        The following steps are performed. We first prepare a state $\ket{\psi}=\matU_{S}\ket{0}$ where $\matU_{S}\sim\matPi$.
        Then, we use Theorem \ref{theorem:approximate_ground_energy} with inputs $\ket{\psi}$, $\matU_{\matHdagH}$, $\epsilon=\epsilon_{\sigma}/2$, and $\delta = 1/4-\delta_{\rm osp}/2\in(0,1/4]$.
        The algorithm returns a bit-string $\widetilde \sigma^2\in[0,1]$ which satisfies $|\widetilde \sigma^2-\sigma_N(\matHdagHtilde)|\leq \epsilon_{\sigma}/2$ with probability at least: 
        \begin{align*}
            1-\delta_{\rm osp}-\delta=3/4-\delta_{\rm osp}/2\in(1/2,3/4].
        \end{align*} 
        Here $\matHdagHtilde$ is the block-encoded version of $\matH^\dagger\matH$, which satisfies $\|\matHdagHtilde-a^2\matHdagH\|\leq 2a\epsilon_{H}$. 
        
        From the stability of singular values, which is a consequence of Weyl's inequality, we conclude that $|\widetilde \sigma^2-\sigma_N^2(\matH)|\leq 2a\epsilon_{H}+\epsilon_{\sigma}/2\leq \epsilon_{\sigma}$, due to the assumption that $\epsilon_{\rm enc}\leq \epsilon_{\sigma}/4a$. From Theorem \ref{theorem:approximate_ground_energy}, the cost of such a run is bounded by
        \begin{itemize}
            \item $O\left(
                \frac{a^2}{\gamma\epsilon_{\sigma}}
                \log(\frac{1}{\gamma})
                \log(\frac{a}{\epsilon_{\sigma}})
                \log(
                    \frac{
                    \log(a/\epsilon_{\sigma})
                    }
                    {1-2\delta_{\rm osp}}
                )
            \right)$
            queries to $\matU_{\matH^\dagger\matH}$,
            \item $O\left(
                \frac{1}{\gamma}
                \log(\frac{a}{\epsilon_{\sigma}})
                \log(
                    \frac{
                    \log(a/\epsilon_{\sigma})
                    }
                    {1-2\delta_{\rm osp}}
                )
            \right)$
            queries to $\matU_{S}$,
            \item $O\left(\log(N) + b + \log(\frac{1}{\gamma})\right)$ qubits,
            \item $O\left(
                \frac{a^2b}{\gamma\epsilon_{\sigma}}
                \log(\frac{1}{\gamma})
                \log(\frac{a}{\epsilon_{\sigma}})
                \log(
                    \frac{
                    \log(a/\epsilon_{\sigma})
                    }
                    {1-2\delta_{\rm osp}}
                )
            \right)$ other elementary gates.
        \end{itemize}

To boost the success probability we can perform multiple runs and keep the median (often referred to as the ``median trick''). 
Note that the returned value $\widetilde \sigma^2$ is always in $[0,1]$, regardless whether the algorithm succeeds or not. Now, assume that we execute $r$ runs, and consider the following random variables for each run:
\begin{align*}
    X_i=\begin{cases}
        1, \text{ if } |\widetilde\sigma^2-\sigma_N^2|\leq \epsilon_{\sigma},\\
        0, \text{else},
    \end{cases}
\end{align*}
where $i=1,\ldots, r$. $X_i$ are independent Bernoulli variables which succeed with probability $p$, where $p\geq 3/4-\delta_{\rm osp}/2$. Let $X=\sum_{i=1}^rX_i$. We need to bound the probability that $X>r/2$. It holds that $\mathbb{E}[X]=rp$. Since, by assumption, $\delta_{\rm osp}<1/2,$ we can use a multiplicative Chernoff bound (see e.g. \cite{bandeira2020mathematics}), which gives $
    \Pr[X<(1-\eta)rp] < \exp(-\eta rp).$
For $\eta = 1-\frac{1}{2p}$, we have that 
\begin{align*}
    \Pr[X\leq r/2] < \exp(-r(p-1/2)) \leq 
    \exp(
        -r(1-2\delta_{\rm osp})/4).
\end{align*}
Setting $r=O\left(\frac{\log(1/\delta)}{1-2\delta_{\rm osp}}\right)$ we obtain the desired success probability $1-\delta$.
\end{proof}
\end{lemma}

\subsection{Counting eigenvalues}
In this section we describe the second subroutine of our algorithm, which is used to count the eigenvalues of a matrix that are smaller than a given threshold. This is a quantum variant of a standard subroutine in spectral divide-and-conquer eigensolvers \cite{demmel1997applied,bai1997inverse,nakatsukasa2013stable,banks2023pseudospectral,demmel2024generalized,sobczyk2024invariant}, often referred to as ``NEGCOUNT.'' Formally, we have the following lemma.
\begin{lemma}[$\QCOUNT$]
    \label{lemma:qcount} Let $\matA$ be a Hermitian matrix with $\|\matA\|\leq 1$, and a promise that $\sigma_{\min}(\matA)\geq \Delta$, for some $\Delta>0$. Given $\matU_{\matA}$, an $(a,b,\epsilon_{\rm enc})$-QUBE for $\matA$, if $\epsilon_{\rm enc}\leq \tfrac{\Delta^2}{32a^2C_{\rm sgn}}N$ for some absolute constant $C_{\rm sgn}$, 
    then there exists a quantum algorithm which returns $z\gets \QCOUNT(\matU_{\matA},\Delta,\delta)$, the true number of eigenvalues of $\matA$ that are smaller than zero, with probability at least $1-\delta$,  with a cost:
    \begin{enumerate}
            \item Queries to $\matU_{\matA}$: $O\left(
                 \tfrac{a^2N}{\Delta}\log\left(N\right)
                    \log(\tfrac{1}{\delta})
            \right)$,
            \item Number of qubits: $O\left(\log(N) + b\right)$,
            \item Other elementary gates: $O\left(
                 b\tfrac{a^2N}{\Delta}\log\left(N\right)
                    \log(\tfrac{1}{\delta})
            \right)$,
            \item Total cost: $O\left(\left(\qcost(\matU_{\matA};\epsilon_{\rm enc})+b\right)
                \cdot \tfrac{a^2N}{\Delta}\log\left(N\right)
                    \log(\tfrac{1}{\delta})
            \right)$.
    \end{enumerate}
    \begin{proof}
        The following steps achieve the result. 
        \begin{enumerate}
           
            \item We use Corollary \ref{corollary:sign_block_encoding} to prepare a block-encoding for {$\sgn(\matA)$}. 
            As long as $\epsilon_{\rm enc} \leq \epsilon_{\rm sgn}\tfrac{\Delta^2}{C_{\rm sgn}}a^2$ for some absolute constant $C_{\rm sgn}$, the corollary returns an $(1,b+2,\epsilon_{\rm sgn})$-QUBE $\matU_{\rm sgn}$ for {$\sgn(\matA)$} with cost \begin{align*}
            \qcost(\matU_{\rm sgn};\Delta,\epsilon_{\rm enc})
            &=
            O\lpar
                (\qcost(\matU_{\matA};\epsilon_{\rm enc})
                +b)\cdot \tfrac{a}{\Delta}\log(\tfrac{1}{\epsilon_{\rm sgn}})
            \rpar 
            .
            \end{align*}
            \item Thereafter, Corollary \ref{corollary:quantum_trace_p} is applied to compute the trace of $\sgn(\matA)$ with accuracy $\epsilon_{\rm tr}$, which takes time 
            \begin{align}
                O\lpar
                    \qcost(\matU_{\rm sgn};\Delta,\epsilon_{\rm enc})  
                    \tfrac{aN}{\epsilon_{\rm tr}}\log(\tfrac{1}{\delta})
                \rpar
                &=
                O\left(
                    \left(\qcost(\matU_{\matA};\epsilon_{\rm enc})
                +b\right) \tfrac{a}{\Delta}\log(\tfrac{1}{\epsilon_{\rm sgn}})
                    \tfrac{aN}{\epsilon_{\rm tr}}\log(\tfrac{1}{\delta})
                \right)
                \nonumber
                \\
                &=
                O\left(
                    \left(\qcost(\matU_{\matA};\epsilon_{\rm enc})+b\right)
                    \tfrac{a^2N}{\Delta\epsilon_{\rm tr}}\log\left(\tfrac{1}{\epsilon_{\rm sgn}}\right)
                    \log(\tfrac{1}{\delta})
                \right).
            \end{align}
        \item We now set $\epsilon_{\rm sgn}=1/(8N)$, $\epsilon_{\rm tr}=1/4$, and $\epsilon_{\rm enc}=\epsilon_{\rm sgn}\tfrac{\Delta^2}{C_{\rm sgn}a^2}=\tfrac{\Delta^2}{8a^2CN}$. Note that this value of $\epsilon_{\rm enc}$ satisfies all the requirements of the previous steps. Plugging-in the parameters, the total complexity becomes
        \begin{align*}
            O\left(
                \left(\qcost(\matU_{\matA};\epsilon_{\rm enc})+b\right)
                \cdot \tfrac{a^2N}{\Delta}\log\left(N\right)
                    \log(\tfrac{1}{\delta})
            \right).
        \end{align*}
        \end{enumerate}
        It remains to verify the correctness. Since $\epsilon_{\rm enc}<\Delta/2$, the number of eigenvalues of $\matA$ 
        (and also $\matAtilde$) that are smaller than zero,
        is equal to $z=\frac{1}{2}\tr(\matI-\sgn({\matAtilde})) = \frac{N}{2}-\frac{1}{2}\tr(\sgn(\matAtilde))= \frac{N}{2}-\frac{1}{2}\tr(\sgn(\matA))$.
        
        From Corollary \ref{corollary:quantum_trace_p}, the above steps return a value $\widetilde t$ which satisfies $\labs \widetilde t-\tr(\sgn(\matA))\rabs
        \leq  
        1/4$. 
        This means that the value $\widetilde z=N/2-\widetilde t/2$ satisfies $|z-\widetilde z| \leq 1/8$. Therefore, we can round $\widetilde z$ to the nearest integer to recover the true $z$. 
    \end{proof}    
\end{lemma}

\section{Main algorithms}
\label{section:main_algorithm}
We now have all the required tools to describe our main algorithms for the $k$-th spectral gap and midpoint, namely, $\gap_k=\lambda_{k}-\lambda_{k+1}$ and $\mu_k=\frac{\lambda_k+\lambda_{k+1}}{2}$. 
Two algorithms are described. The first one, Algorithm \ref{algorithm:qgap_const} takes as input a Hermitian matrix $\matH$ (scaled to have $\|\matH\|\leq 1/2$), an integer $k\in[N-1]$, and a failure probability parameter $\delta$. It returns constant factor approximations to $\gap_k$ and $\mu_k$. 

The second Algorithm \ref{algorithm:qeig}, takes as input $\matH$, $k$, $\delta$, and an accuracy parameter $\epsilon$. It first uses Algorithm \ref{algorithm:qgap_const} as a subroutine to compute a coarse (constant factor) approximation to $\gap_k$ and $\mu_k$, and then uses these approximations to refine the accuracy of the solution to the given $\epsilon$. As a side-result, it also returns $\epsilon$-approximations to the actual eigenvalues $\lambda_k$ and $\lambda_{k+1}$ around the gap.

The complexity of these algorithms depends on the value of $\gap_k$. However, we highlight that no prior knowledge or lower bound on $\gap_k$ is required. The only required parameters are the ones given as input to the algorithms. All other parameters that are involved in the complexity are computed/approximated on-the-fly, if required.
\begin{lemma}[Alg. \ref{algorithm:qgap_const}, $\QGAPconst$]
    \label{lemma:alg_qgap_const}
    Let $\matH$ be an $N\times N$ Hermitian matrix with eigenvalues
    \begin{align*}
        1/2\geq \lambda_1 \geq \lambda_2 \geq \ldots \geq \lambda_{N} \geq -1/2.
    \end{align*}
    Let:
    \begin{itemize}
        \item $\qcost(\matU_{\matH};\epsilon_{\rm enc})$ be      the cost of the unitary $\matU_{\matH}$ which forms an $(a,b,\epsilon_{\rm enc})$-QUBE of $\matH$.
        \item $\matPi$ be a $(\gamma,\delta_{\rm osp})$-OSP distribution, where $\matU_S\sim \matPi$ has cost $\qcost(\matU_{S})$. 
    \end{itemize} 
    Then, given $\matH$
    and an integer $k\in[N-1]$, Algorithm \ref{algorithm:qgap_const} ($\QGAPconst(\matH, k, \delta)$) computes an approximate midpoint $\widetilde\mu_k$ and gap $\widetilde \gap_k$, such that:
    \begin{align*}
        \labs \widetilde \mu_k - \mu_k \rabs \leq \frac{7}{16}\gap_k,
        \qquad
        \widetilde \gap_k\in[\gap_k/2,4\gap_k].
    \end{align*}
   The algorithm succeeds with probability at least $1-\delta$. It makes at most: 
   \begin{enumerate}
       \item Queries to unitaries with gate complexity  $O\left(\qcost(\matU_{\matH};\epsilon_{\rm enc})\right)$, where $\epsilon_{\rm enc}\in \Theta\left(\frac{\gap_k^{2}}{a}\right)$:
       \begin{align*}
           Q_1:=O\left(
                \frac{a^2}{\gamma\gap_k^{2}(1-2\delta_{\rm osp})}
                \log(\frac{1}{\gamma})
                \log(\frac{a}{\gap_k})
                \log(
                    \frac{
                    \log(a/\gap_k)
                    }
                    {1-2\delta_{\rm osp}}
                )
                \log(\frac{1}{\delta})
                \log(\frac{1}{\gap_k})
            \right),
       \end{align*}
            \item Queries to $\matU_{S}$:
            \begin{align*}
                O\left(
                \frac{1}{\gamma(1-2\delta_{\rm osp})}
                \log(\frac{a}{\gap_k})
                \log(
                    \frac{
                    \log(a/\gap_k)
                    }
                    {1-2\delta_{\rm osp}}
                )
                \log(\frac{1}{\delta})
                \log(\frac{1}{\gap_k})
                \right),
            \end{align*}
            \item Queries to unitaries with gate complexity $O\left(\qcost(\matU_{\matH};\epsilon'_{\rm enc})\right)$, where $\epsilon_{\rm enc}'
            \in
            \Theta\left(
                \frac{\gap_k^{2}}{a^2N}
            \right)$:
            \begin{align*}
                Q_2:=O\left(
                    \frac{a^2N}{\gap_k}
                    \log\left(N\right)
                    \log(\frac{1}{\gap_k\delta})
                    \log(\frac{1}{\gap_k})
                \right),
            \end{align*}
            \item Number of qubits: $O(\log(N)+b+\log(1/\gamma)),$
            \item Other elementary gates: $O\left(b\cdot(Q_1+Q_2)\right)$.
            
   \end{enumerate}
    \begin{proof} 
        The algorithm executes the following steps. Initially, it sets $j=2$, and $\gap=1/2$. At each iteration $j=2,\ldots$, the goal is to search for all gaps that are larger than $\gap^j=\frac{1}{2^j}$.
        
        In the first step of each iteration, a uniform grid with points $h_i=-\frac{1}{2}+i\frac{\gap^j}{2}$, $i=0,\ldots,\frac{2}{\gap^j}$, such that $|h_i-h_{i+1}|=\gap^j/2$. If there exists \textit{any} gap with width at least $\gap^j$, then there exists at least one point $h_i$ such that $\sigma_{\min}(\matH-h_i)\geq \gap^j/4$. 
        For each $h_i$ in the grid, the algorithm proceeds as follows: 
        \begin{enumerate}
            \item Use Lemma~\ref{lemma:sgn_new} to create an $(a+|h_i|,b+1,\epsilon_{\rm enc})$-QUBE $\matU_{i}$ of $\matH-h_i$, where we set $\epsilon_{\rm enc}:=(\frac{\gap^{j}}{8})^2\frac{1}{4a}$.
            The cost of $\matU_i$ is $\qcost(\matU_{\matH};\epsilon_{\rm enc}) + O(1)$.

            \item Execute $\QSMIN$ with parameters $\epsilon_{\sigma}=\frac{\gap^{2j}}{64}$ and $\delta_{\sigma}=\frac{\delta_j}{2}$ to test if $\sigma_{\min}(\matH-h_i)$ satisfies the desired bound.            
        \end{enumerate}

        The requirement $\epsilon_{\rm enc}\leq \frac{\epsilon_{\sigma}}{4a}$ of Lemma 
        \ref{lemma:qsmin} is satisfied, and therefore $\QSMIN$ returns a value $\widetilde\sigma_i^2$ which satisfies 
        \begin{align*}
            \Pr\left[
                \left|
                    \widetilde\sigma_i^2-\sigma_{\min}^2(\matH-h_i)
                \right| > \frac{\gap^{2j}}{64}
            \right] 
            \leq \frac{\delta_j}{2}.
        \end{align*}
        This means that, for each $h_i$, the algorithm can test if the computed $\widetilde\sigma_i^2$ satisfies $\widetilde \sigma_i^2\geq \gap^{2j}/32$, and,  if this holds for any $i$, then it is ensured that $\sigma^2_{\min}(\matH-h_i)\geq \widetilde\sigma^2_i-\epsilon_{\sigma}> \frac{ \gap^{2j}}{64}$, or, in other words,  $\sigma_{\min}(\matH-h_i) > \frac{ \gap^{j}}{8}$.
        
        Now, this lower bound allows us to count the eigenvalues smaller than $h_i$ efficiently using $\QCOUNT$: 
        \begin{enumerate}
        \setcounter{enumi}{2}
            \item Use Lemma~\ref{lemma:sgn_new} to construct an
        $(a+|h_i|,b+1,\epsilon'_{\rm enc})$-QUBE $\matU'_{i}$ for $\matH-h_i$ with $\epsilon'_{\rm enc}:= \tfrac{\Delta^{2j}}{64\cdot 32C_{\rm sgn}a^2N}$.
        Here $C_{\rm sgn}$ is the constant from Lemma~\ref{lemma:qcount} (note that this value of $\epsilon'_{\rm enc}$ satisfies the requirement of the lemma).
        The cost of $\matU'_i$ is $\qcost(\matU_{\matH};\epsilon'_{\rm enc}) + O(1)$.
            \item By executing $\QCOUNT(\matU'_{i},\frac{\gap^{j}}{8},\frac{\delta_j}{2})$, we obtain $z_i$, the true number of eigenvalues that are smaller that $h_i$, with probability at least $1-\delta_j/2$.
        \end{enumerate}

        We can now calculate the total costs for each iteration. By replacing $\epsilon_{\sigma} = \frac{\gap^{2j}}{64}$ and $\delta_\sigma =\delta_j/2=\frac{\delta\gap^j}{2^{j+2}}$ in Lemma \ref{lemma:qsmin}, each call to $\QSMIN$ costs:
         \begin{enumerate}
            \item Queries to $\matU_{i}$: \begin{align*}
                O\left(
                \frac{a^2}{\gamma\gap^{2j}}
                \log(\frac{1}{\gamma})
                \log(\frac{a}{\gap^{2j}})
                \log(
                    \frac{
                    \log(a/\gap^{2j})
                    }
                    {1-2\delta_{\rm osp}}
                )
                \frac{\log(2^{j+2}/(\gap^j\delta))}{1-2\delta_{\rm osp}}
            \right).
            \end{align*}
            \item Queries to random state-preparation circuits $\matU_{S}$:
            \begin{align*}
            O\left(
                \frac{1}{\gamma}
                \log(\frac{a}{\gap^{2j}})
                \log(
                    \frac{
                    \log(a/\gap^{2j})
                    }
                    {1-2\delta_{\rm osp}}
                )
                \frac{\log(2^{j+2}/(\gap^j\delta))}{1-2\delta_{\rm osp}}
            \right).
            \end{align*}
            \item Number of qubits:
            \begin{align*}
            O\left(
                \log(N) + b + \log(\frac{1}{\gamma})
            \right).
            \end{align*}
            \item Other elementary gates:
            \begin{align*}
            O\left(
                \frac{a^2b}{\gamma\gap^{2j}}
                \log(\frac{1}{\gamma})
                \log(\frac{a}{\gap^{2j}})
                \log(
                    \frac{
                    \log(a/\gap^{2j})
                    }
                    {1-2\delta_{\rm osp}}
                )
                \frac{\log(2^{j+2}/(\gap^j\delta))}{1-2\delta_{\rm osp}}
            \right).
            \end{align*}
        \end{enumerate}
        From Lemma \ref{lemma:qcount}, each call to $ \QCOUNT\left(\matU'_i,\frac{\Delta^j}{8}, \frac{\delta \Delta^j}{2^{j+1}}\right)$ costs: 
        \begin{enumerate}
            \item Queries to $\matU'_i$: $O\left(\tfrac{a^2N}{\gap^{j}}\log\left(N\right)
                    \log(\tfrac{2^j}{\gap^j\delta})\right)$,
            \item Number of qubits: $O(\log(N) + b)$,
            \item Other elementary gates: $O\left(b\tfrac{a^2N}{\gap^{j}}\log\left(N\right)
                    \log(\tfrac{2^j}{\gap^j\delta})\right)$,
            \item Total cost: $O\lpar
     \left(\qcost(\matU_{\matH};\epsilon'_{\rm enc}) + b\right)\cdot \tfrac{a^2N}{\gap^{j}}\log\left(N\right)
                    \log(\tfrac{2^j}{\gap^j\delta})
    \rpar$.
        \end{enumerate}
        
        The number of iterations can be bounded as follows. First, note that a sufficient termination criterion is $\gap^j\leq\gap_k$. If this holds, there exists at least one point $h_i$ in the grid that satisfies $\sigma_{\min}(\matH-h_i)\geq \frac{\Delta_k}{4}>\frac{\Delta^j}{4}$. 
        Subsequently, the approximate singular value returned by $\QSMIN$ will satisfy $\widetilde\sigma_i^2\geq \sigma_i^2-\gap^{2j}/64 > \frac{1.5\gap^{2j}}{32}$, which means that the algorithm will certainly execute $\QCOUNT$ for that point $h_i$. 
        Ultimately, $\QCOUNT$ will verify that $h_i$ is  inside the correct gap, and the algorithm will terminate. 
        
        We can obtain a necessary termination criterion $  \frac{\gap^j}{4}\leq\gap_k$ in a similar way. 
        To see this, assume that the algorithm terminated after finding a point that satisfies $\widetilde\sigma_i^2>\frac{\gap^{2j}}{32}$. This implies that $\sigma_i^2>\frac{\gap^{2j}}{64} \implies \sigma_i>\frac{\gap^{j}}{8}$. 
        But since $h_i$ is inside the correct $k$-th gap, it means that $\gap_k\geq 2\sigma_i\geq \frac{\gap^j}{4}$, as initially claimed.

        Since we are halving $\gap^j$ in every iteration, we conclude that the algorithm terminates after  $m:=\lceil\log(1/\gap_k)\rceil+C_{\rm gap}$ iterations, where $C_{\rm gap}\in[0,2]$. This means that $\gap^m=(1/2)^m\in\Theta(\gap_k)$.
        The most expensive iteration is the last one, where the corresponding block-encodings have costs:
        \begin{itemize}
            \item $Q:=\qcost(\matU_i;\epsilon_{\rm enc})\in \qcost(\matU_{\matH},\epsilon_{\rm enc})+O(1)$, where \begin{align*}
            \epsilon_{\rm enc}= \left(
                \frac{\gap^{\lceil\log(1/\gap_k)\rceil
                +
                C_{\rm gap}}}{8}
            \right)^2\frac{1}{4a} \in \Theta\left(
            \frac{\gap_k^{2}}{a}
            \right),
            \end{align*}
            \item $Q':=\qcost(\matU'_i;\epsilon'_{\rm enc})\in \qcost(\matU'_{\matH},\epsilon'_{\rm enc})+O(1)$, where 
            \begin{align*}
                \epsilon_{\rm enc}'
            =
            \frac{\gap^{2\lceil\log(1/\gap_k)\rceil
                +
                2C_{\rm gap}}}{64\cdot 32a^2C_{\rm sgn}N}
            \in
            \Theta\left(
                \frac{\gap_k^{2}}{a^2N}
            \right).
            \end{align*}
        \end{itemize}
        The total cost of the last iteration is therefore:
        \begin{align*}
            &O\Bigg(
            \lbrac
                \left(Q+b\right)\cdot 
                \tfrac{a^2}{\gamma\gap^{2m}}
                \log(\tfrac{1}{\gamma})
                +
                \qcost(\matU_{S})
                \tfrac{1}{\gamma}
            \rbrac
                \log(\tfrac{a}{\gap^{2m}})
                \log(
                    \tfrac{
                    \log(a/\gap^{2m})
                    }
                    {1-2\delta_{\rm osp}}
                )
                \tfrac{\log(2^{m+2}/(\gap^m\delta))}{1-2\delta_{\rm osp}}
            \\ 
            &\qquad+ 
            \left(Q'+b\right)\cdot 
            \tfrac{a^2N}{\gap^{m}/2}
            \log\left(N\right)
            \log(\tfrac{2^m}{\gap^m\delta})
        \Bigg)
        \end{align*}
        Let us bound each additive term separately. By replacing $\gap=1/2$ we have:
        \begin{align}
            &O\left(
            \lbrac
                \left(Q+b\right)\cdot
                \tfrac{a^2}{\gamma\gap^{2m}}
                \log(\tfrac{1}{\gamma})
                +
                \qcost(\matU_{S})
                \tfrac{1}{\gamma}
            \rbrac
                \log(\tfrac{a}{\gap^{2m}})
                \log(
                    \tfrac{
                    \log(a/\gap^{2m})
                    }
                    {1-2\delta_{\rm osp}}
                )
                \tfrac{\log(2^{m+2}/(\gap^m\delta))}{1-2\delta_{\rm osp}}
            \right)\nonumber
            \\
            &\qquad=
            O\left(
            \lbrac
                \left(Q+b\right)\cdot 
                \tfrac{a^2}{\gamma\gap_k^{2}}
                \log(\tfrac{1}{\gamma})
                +
                \qcost(\matU_{S})
                \tfrac{1}{\gamma}
            \rbrac
                \log(\tfrac{a}{\gap_k^{2}})
                \log(
                    \tfrac{
                    \log(a/\gap_k^{2})
                    }
                    {1-2\delta_{\rm osp}}
                )
                \tfrac{\log(4/\delta)}{1-2\delta_{\rm osp}}
            \right).
        \end{align}
        For the second term we have
        \begin{align}
        O\left(
            \left(Q'+b\right)\cdot  
            \tfrac{a^2N}{\gap^{m}}
            \log\left(N\right)
            \log(\tfrac{2^m}{\gap^m\delta})
        \right)=
        O\left(
            \left(Q'+b\right)\cdot 
            \tfrac{a^2N}{\gap_k}
            \log\left(N\right)
            \log(\tfrac{1}{\gap_k^{2}\delta})
        \right).
        \end{align}
        After simplifying some constants, the final total cost is bounded by the sum of the following terms:
        \begin{itemize}
            \item $O\left(
                 \left(\qcost(\matU_{i};\epsilon_{\rm enc})+b\right)
                \frac{a^2}{\gamma\gap_k^{2}}
                \log(\frac{1}{\gamma})
                \log(\frac{a}{\gap_k})
                \log(
                    \frac{
                    \log(a/\gap_k)
                    }
                    {1-2\delta_{\rm osp}}
                )
                \frac{\log(1/\delta)}{1-2\delta_{\rm osp}}
                \log(\frac{1}{\gap_k})
            \right)$,
            \item $
            O\left(
            \qcost(\matU_{S})
                \frac{1}{\gamma}
                \log(\frac{a}{\gap_k})
                \log(
                    \frac{
                    \log(a/\gap_k)
                    }
                    {1-2\delta_{\rm osp}}
                )
                \frac{\log(1/\delta)}{1-2\delta_{\rm osp}}
                \log(\frac{1}{\gap_k})
                \right)$,
            \item $
                O\left(
                    \left(\qcost(\matU'_{i};\epsilon'_{\rm enc})+b\right) 
                    \frac{a^2N}{\gap_k}                    \log\left(N\right)
                    \log(\frac{1}{\gap_k\delta})
                    \log(\frac{1}{\gap_k})
                \right),
            $
        \end{itemize}
        where $\qcost(\matU_{i};\epsilon_{\rm enc})$ and $\qcost(\matU'_{i};\epsilon'_{\rm enc})$ depend on the values of $\epsilon_{\rm enc}$ and $\epsilon_{\rm enc}'$, respectively, in the final iteration, and the trailing $\log(\tfrac{1}{\gap_k})$ factor is the number of iterations.

        Next, we argue about the correctness of the algorithm and the approximation guarantees of the final results. Once we find some $h_i$, for which the result of $\QCOUNT$ satisfies $z_i=k$, we know that the point $h_i$ is inside the correct gap $\gap_k$. The algorithm then returns $\widetilde\mu_k\gets h_i$ and $\widetilde\gap_k\gets \Delta^j$, which satisfy the following bounds. In the above, we already argued that $\gap_k\geq \gap^j/4$ is necessary for the algorithm to terminate. But we can also see that $\gap^j\geq\gap_k/2$ holds as well. Indeed, the algorithm cannot terminate when $\gap^j< \gap_k/2$, otherwise it would have terminated in the previous step.
        The combination of these two facts implies that the returned $\gap^m$ satisfies:
        \begin{align}
            \gap_k/2 \leq \gap^m \leq 4\gap_k.
        \end{align}
        Combining this with the termination criterion of the algorithm, we also conclude that the returned midpoint $\widetilde\mu_k$ satisfies:
        \begingroup
        \allowdisplaybreaks
        \begin{align} 
        \widetilde\mu_k&\in[\lambda_{k+1}+\gap^m/8,\lambda_{k}-\gap^m/8]
        \nonumber\\
        &\in
        [\lambda_{k+1}+\gap_k/16,\lambda_{k}-\gap_k/16]
        \nonumber\\
        &\in
        [\mu_k-7\gap_k/16,\mu_{k}+7\gap_k/16],
        \end{align}
        \endgroup
        where in the last we replaced $\lambda_{k+1}=\mu_k-\frac{\gap_k}{2}$ and $\lambda_{k}=\mu_k+\frac{\gap_k}{2}$.
        
        Finally, we bound the success probability. 
        There are a total of at most $l_j:=1+2/\gap^j$ calls to $\QCOUNT$ and $\QSMIN$ in each iteration $j$. Each call  fails with probability at most $p_j:=\delta_j/2$. 
        Thus, the total success probability for iteration $j$ is at least 
        \begin{align*}
            1-2\cdot l_j p_j
            =
            1-2\cdot\frac{\delta_j}{2}\lpar1+\frac{2}{\gap^j}\rpar
            =
            1-\frac{\delta\gap^j}{2^{j+1}}\lpar1+\frac{2}{\gap^j}\rpar
            \geq
            1-\frac{\delta}{2^{j}}.
        \end{align*}
        Taking a union bound over all iterations, we have
        \begin{align*}
            p_{success} \geq 1-\sum_{j=1}^{m} (\delta/2^j) \geq 1-\delta\sum_{j=1}^{\infty}2^{-j} = 1-\delta.
        \end{align*}
    \end{proof}
\end{lemma}
\begin{algorithm}[hbt]
\begingroup
\footnotesize
\setstretch{1.5}  
\begin{algorithmic}[1]
\Algorithm{$\widehat\mu_k,\widehat\gap_k\gets\QGAPconst(\matH,k,\delta)$}
    \Statex \textbf{Input:}{ Hermitian matrix $\matH\in\mathbb{C}^{N\times N}$ with $\|\matH\|\leq 1/2$, integer $k\in[N-1]$, and failure probability parameter $\delta\in(0,1/2)$. Optionally, a threshold $\gap_{\min}$ which is used to terminate the algorithm early.}
    \Statex \textbf{Output:}{ Approximate gap $\widetilde \gap_k$ and midpoint $\widetilde \mu_k$, satisfying the guarantees of Lemma \ref{lemma:alg_qgap_const} with probability $1-\delta$}.
    \State Set $j=2$, $\Delta\gets 1/2.$
    \While{$\gap^j>\gap_{\min}$}
        \State Construct a grid in $\left[-\tfrac{1}{2},\tfrac{1}{2}\right]$, with points $h_i=-\frac{1}{2}+i\frac{\gap^j}{2}$, $i=0,\ldots,\frac{2}{\gap^j}$
        \State $\delta_j\gets \frac{\delta\gap^j}{2^{j+1}}$ \Comment{Failure probability for iteration $j$}
        
        \State $\epsilon_{\rm enc}\gets (\frac{\gap^{j}}{8})^2\frac{1}{4a}$
        \State $\epsilon'_{\rm enc}\gets
        \epsilon_{\rm enc}\frac{1}{8C_{\rm sgn}aN}
        =
        \tfrac{\gap^{2j}}{64\cdot 32a^2C_{\rm sgn}N}$
        \State $\matU_{\matH}\gets \QENC_{a,b}\left(
                \matH,\epsilon_{\rm enc}
            \right)$
        \State $\matU'_{\matH}\gets \QENC_{a,b}\left(
                \matH,\epsilon'_{\rm enc}
            \right)$
        \For{all $h_i$, $i=0,\ldots, \tfrac{2}{\gap^j}$}
            \State $\matU_{i}\gets \QSHIFT\left(
                \matU_{\matH},h_i,a
            \right)$
            \State  $\widetilde \sigma_i^2\gets\QSMIN
            \left(
            \matU_{i},\tfrac{\gap^{2j}}{64},\tfrac{\delta_j}{2}
            \right)$
            \If{$\widetilde \sigma_i^2\geq \tfrac{\gap^{2j}}{32}$}
                \State $\matU'_{i}\gets \QSHIFT\left(
                    \matU'_{\matH},h_i,a
                \right)$
                \State  $z_i\gets \QCOUNT\left(\matU'_{i},\tfrac{\gap^j}{8},\tfrac{\delta_j}{2}\right)$
            \EndIf
            \If{$z_i=N-k$}
                    \State \Return $\widetilde \gap_k\gets\gap^j$, and $\widetilde\mu_k\gets h_i$
            \EndIf
        \EndFor
        \State $j\gets j+1$
    \EndWhile
    \State \Return NULL.
\end{algorithmic}
\endgroup
\caption{Constant factor approximation to the $k$-th spectral gap and midpoint.}
\label{algorithm:qgap_const}
\end{algorithm}
With some additional steps the solution accuracy can be increased to any $\epsilon>0$.
\begin{theorem}
    \label{theorem:alg_qeig}
    Let $\matH$ be an $N\times N$ Hermitian matrix, $\|\matH\|\leq 1/2$, $k\in[N-1]$ be an integer, and $\epsilon,\delta\in(0,1)$ an accuracy and failure probability parameter, respectively. Let 
    \begin{itemize}
        \item $\qcost(\matU_{\matH};\epsilon_{\rm enc})$ denote the cost of the unitary $\matU_{\matH}$ which forms an $(a,b,\epsilon_{\rm enc})$-QUBE of $\matH$.
        \item $\matPi$ be a $(\gamma,\delta_{\rm osp})$-OSP distribution. 
    \end{itemize} 
    Algorithm \ref{algorithm:qeig} returns $\widetilde\lambda_k, \widetilde\lambda_{k+1}, \widetilde\mu_k,\widetilde\gap_k$ that satisfy:
    \begingroup
    \allowdisplaybreaks
    \begin{align*}
        \widetilde\lambda_k&\in[\lambda_k-\epsilon\gap_k/2,\lambda_k],
        &
        \widetilde\lambda_{k+1}&\in[\lambda_{k+1},\lambda_{k+1}+\epsilon\gap_k/2]\\
        \widetilde\mu_k &\in \mu_k\pm \epsilon\gap_k/2,
        &
        \widetilde\gap_k&\in(1\pm\epsilon)\gap_k.
    \end{align*}
    \endgroup
    In addition to the costs induced by calling Algorithm \ref{algorithm:qgap_const} in the first step, it requires 
    \begin{enumerate}
            \item Queries to unitaries with complexity $O(\qcost(\matU_{\matH};\epsilon_{\rm enc}))$, where $\epsilon_{\rm enc}\in\Theta(\gap_k^2/a)$:
            \begin{align*}
                Q_1:=O\left(
                \frac{a^2}{\gamma\epsilon^2\gap_k^2}
                \log(\frac{1}{\gamma})
                \log(\frac{a}{\epsilon\gap_k})
                \log(
                    \frac{
                    \log(a/(\epsilon\gap_k))
                    }
                    {1-2\delta_{\rm osp}}
                )
                \frac{\log(1/(\epsilon\delta))}{1-2\delta_{\rm osp}}
            \right),
            \end{align*}
            \item Queries to OSP circuits $\matU_{S}\sim \matPi$:
            \begin{align*}
                O\left(
                \frac{1}{\gamma}
                \log(\frac{a}{\epsilon\gap_k})
                \log(
                    \frac{
                    \log(a/(\epsilon\gap_k))
                    }
                    {1-2\delta_{\rm osp}}
                )
                \frac{\log(1/(\epsilon\delta))}{1-2\delta_{\rm osp}}
            \right),
            \end{align*}
            \item Queries to unitaries with complexity $Q_2:=O\left(\qcost(\matU_{\matH};\epsilon'_{\rm enc})\right)$, where $\epsilon'_{\rm enc}\in\Theta(\epsilon^2\gap_k^2/(a^2N))$:
            \begin{align*}
                O\left(
                \tfrac{a^2N}{\epsilon^2\gap_k}\log\left(N\right)
                    \log(\tfrac{1}{\delta})
            \right),
            \end{align*}
            \item Number of qubits: $O\left(\log(N) + b +\log(\frac{1}{\gamma})\right)$,
            \item Other elementary gates: $O\left(b\cdot(Q_1+Q_2)\right)$.
        \end{enumerate}
        If we use Lemma \ref{lemma:oblivious_state_preparation} for $\matPi$, after some simplifications, the total circuit complexity (of both Alg. \ref{algorithm:qgap_const} and \ref{algorithm:qeig}) is upper bounded by
        \begin{align}
        O\left(
            \left[
            N^2
            +
            N\cdot \tfrac{a^2 b\qcost(\matU_{\matH};\epsilon_1)}{\epsilon^2\gap_k^2}
            \right]
            \polylog\left(N,a,\epsilon^{-1},\gap_k^{-1},\delta^{-1}\right)
        \right),
        \label{eq:main_theorem_cosst_with_specific_osp}
        \end{align}
        where $\epsilon_1\in \Theta\left(
        \frac{\epsilon^2\gap_k^2}{a^2N}
        \right)$.
    \begin{proof}
        Let us call $\widehat\mu_k$ and $\widehat\gap_k$ the initial solutions returned by Algorithm \ref{algorithm:qgap_const}, i.e., the solutions that satisfy the constant factor accuracy of Lemma \ref{lemma:alg_qgap_const}. We know that $\widehat\gap_k\in[0.5\gap_k,4\gap_k]$. By setting $\epsilon_k:=\epsilon\widehat\gap_k/4$, it means that $\epsilon_k\leq \epsilon\gap_k$.

        We now approximate both eigenvalues $\lambda_k$ and $\lambda_{k+1}$ to accuracy $\epsilon_k/2$ each as follows. 
        Consider first $\lambda_{k+1}$. 
        We make a grid on the left of $\widehat\mu_k$ with step $\epsilon_k/4$, consisting of the points $l_i=\widehat\mu_k+i\epsilon_k/4$, for $i=-\lceil 2\widehat\gap_k/\epsilon_k\rceil,\ldots, -2, -1$.
        Note that, there exists at least one point $l^*>\lambda_{k+1}$ such that $\epsilon_k/4\leq |\lambda_{k+1}-l^*|<\epsilon_k/2,$ or, equivalently, $\sigma_{\min}(\matH-l^*)\in[\epsilon_k/4,\epsilon_k/2]$.
         
        To find such a point, we first set the parameters:
        \begin{align*}
            \epsilon_{\sigma}\gets \tfrac{3\epsilon_k^2}{64},
            \qquad 
            \epsilon_{\rm enc}\gets\tfrac{\epsilon_{\sigma}}{4a}=\tfrac{3\epsilon_k^2}{256a},
            \qquad
            \delta_{\sigma} \gets \tfrac{\delta}{\lceil2\widehat\gap_k/\epsilon_k\rceil},
            \qquad
            \epsilon'_{\rm enc}\gets \tfrac{\epsilon_k^2}{64\cdot 32a^2C_{\rm sgn}N},
        \end{align*} 
        and prepare two block-encodings:
        \begin{itemize}
            \item $\matU_{\matH}\gets \QENC_{a,b}(\matH,\epsilon_{\rm enc})$,
            \item  
        $\matU'_{\matH}\gets \QENC_{a,b}(\matH,\epsilon'_{\rm enc})$.
        \end{itemize}
        Then we execute the steps 10-18 for each $i$, similar to Algorithm \ref{algorithm:qgap_const}, but with the aforementioned parameters. We can verify that the returned values $\widetilde\lambda_k,\widetilde\lambda_{k+1},\widetilde\mu_k,\widetilde\gap_k,$ satisfy the advertised bounds.

        It remains to bound the complexity. 
        There are a total of $O(\epsilon^{-1})$ points in the two grids, which gives the total number of calls to the subroutines $\QSMIN$ and $\QCOUNT$. 
        At each grid point, we call $\QENC$ with parameter $\epsilon_{\rm enc}=3\epsilon_k^2/256a\in\Theta(\gap_k^2/a)$, followed by $\QSMIN$ with parameters:
        \begin{align*}
            \epsilon_{\sigma}=\tfrac{3\epsilon_k^2}{64}\in\Theta(\epsilon^2\gap_k^2),\quad \text{and} \quad \delta_{\sigma}=\tfrac{\delta}{\lceil2\widehat\gap_k/\epsilon_k\rceil}\in\Theta(\epsilon \delta).
        \end{align*}
        The cost for each $\QSMIN$ call is
        \begin{enumerate}
            \item $O\left(
                \frac{a^2}{\gamma\epsilon^2\gap_k^2}
                \log(\frac{1}{\gamma})
                \log(\frac{a}{\epsilon\gap_k})
                \log(
                    \frac{
                    \log(a/(\epsilon\gap_k))
                    }
                    {1-2\delta_{\rm osp}}
                )
                \frac{\log(1/(\epsilon\delta))}{1-2\delta_{\rm osp}}
            \right)$
            queries to $\matU_{\matH}$,
            \item $O\left(
                \frac{1}{\gamma}
                \log(\frac{a}{\epsilon\gap_k})
                \log(
                    \frac{
                    \log(a/(\epsilon\gap_k))
                    }
                    {1-2\delta_{\rm osp}}
                )
                \frac{\log(1/(\epsilon\delta))}{1-2\delta_{\rm osp}}
            \right)$
            queries to OSP circuits $\matU_{S}\sim \matPi$,
            \item $O\left(\log(N) + b + \log(\frac{1}{\gamma})\right)$ qubits,
            \item $O\left(
                \frac{a^2}{\gamma\epsilon^2\gap_k^2}
                \log(\frac{1}{\gamma})
                \log(\frac{a}{\epsilon\gap_k})
                \log(
                    \frac{
                    \log(a/(\epsilon\gap_k))
                    }
                    {1-2\delta_{\rm osp}}
                )
                \frac{\log(1/(\epsilon\delta))}{1-2\delta_{\rm osp}}
            \right)$ other elementary gates.
        \end{enumerate}
        In the worst case, $\QCOUNT$ will also be called in every iteration. Each call takes as input an $(a,b,\epsilon'_{\rm enc})$–QUBE, accuracy $\epsilon'_{\rm enc}=\epsilon^2_k/(64a^2C_{\rm sgn}N)\in\Theta(\epsilon^2\gap_k^2/(a^2N)),$ and failure probability $\delta_{\sigma}=\Theta(\epsilon\delta)$.
        Substituting in Lemma \ref{lemma:qcount}, each $\QCOUNT$ call costs:
        \begin{align*}
            O\left(
                \tfrac{a^2N}{\epsilon^2\gap_k}\log\left(N\right)
                    \log(\tfrac{1}{\delta})
            \right)
        \end{align*}
        calls to the block-encoding circuit $\matU'_{\matH}$.        
    \end{proof}
\end{theorem}
\begin{algorithm}[htb]
\begingroup
  \footnotesize
\setstretch{1.5}  
\begin{algorithmic}[1]
\Algorithm{$\lbrac \widetilde \lambda_k,\widetilde\lambda_{k+1},\widetilde\mu_k,\widetilde\gap_k
\rbrac
\gets\QEIG(\matH,k,\delta,\epsilon)$}
     \Statex \textbf{Input:}{ Hermitian matrix $\matH\in\mathbb{C}^{N\times N}$, $\|\matH\|\leq 1/2$, integer $k\in[N-1]$, failure probability parameter $\delta\in(0,1/2)$, accuracy parameter $\epsilon\in(0,1)$.}
    \Statex \textbf{Output:}{ Approximate gap $\widetilde \gap_k$, midpoint $\widetilde \mu_k$, and eigenvalues $\widetilde\lambda_k$, $\widetilde\lambda_{k+1}$, satisfying the guarantees of Theorem \ref{theorem:alg_qeig} with probability at least $1-\delta$}.
    \State $\widehat\mu_k,\widehat\gap_k \gets \QGAPconst(\matH,k,\delta/2)$
    \State $\epsilon_k\gets \epsilon\widehat\gap_k/4$
    \State $\epsilon_{\sigma}\gets \tfrac{3\epsilon_k^2}{64}$ 
    \State $\delta_{\sigma} \gets \tfrac{\delta}{\lceil2\widehat\gap_k/\epsilon_k\rceil}$ \Comment{Failure probability per iteration}
    \State $\epsilon_{\rm enc}\gets\tfrac{\epsilon_{\sigma}}{4a}=\tfrac{3\epsilon_k^2}{256a}$
    \State $\epsilon'_{\rm enc}\gets \tfrac{\epsilon_k^2}{64\cdot 32a^2CN}$
    \State $\matU_{\matH}\gets \QENC_{a,b}(\matH,\epsilon_{\rm enc})$
    \State $\matU'_{\matH}\gets \QENC_{a,b}(\matH,\epsilon'_{\rm enc})$
    \State $M\gets \lceil 2\widehat\gap_k/\epsilon_k\rceil$
    \For{$i=-M,-M+1,\ldots,M-1, M$}
    \State $l_i\gets\widehat\mu_k+i\epsilon_k/4$ 
    \State $\matU_i \gets \QSHIFT(\matU_{\matH},l_i,a)$
            \State $\widetilde\sigma_i^2\gets \QSMIN(\matU_i,\epsilon_{\sigma},\delta_{\sigma})$
            \If{$\widetilde\sigma_i^2\geq \frac{\epsilon_k^2}{32}$}
                \State $\matU'_i\gets \QSHIFT(\matU'_{\matH},l_i,a)$
                \State $z_i\gets \QCOUNT(\matU'_i,\frac{\epsilon^2_k}{64},\delta_{\sigma})$
            \Else:
                \State $z_i\gets -1$
            \EndIf
        \EndFor
            \State  $\widetilde\lambda_{k}\gets \min_i\{l_i\ |\ z_i=N-k \}$, and  $\widetilde\lambda_{k+1}\gets \max_i\{l_i\ |\ z_i=N-k \}$
            \State
            $\widetilde\mu_k\gets (\widetilde\lambda_k+\widetilde\lambda_{k+1})/2$
            \State
            $\widetilde\gap_k\gets \widetilde\lambda_k-\widetilde\lambda_{k+1}$
        \State \Return $\widetilde\lambda_k,\widetilde\lambda_{k+1},\widetilde\mu_k,\widetilde\gap_k$
\end{algorithmic}
\endgroup
\caption{Main algorithm for high-accuracy consecutive eigenvalue pair, spectral gap, and midpoint.}
\label{algorithm:qeig}
\end{algorithm}

\section{Complexity in QRAM}
\label{section:qram}
In this section we specialize the complexity analysis in the QRAM model~\cite{giovannetti2008quantum}.
We describe a specific technique to construct block-encodings via the so-called \textit{quantum-accessible} data structures, studied in~ \cite{kerenidis2016quantum,gilyen2019quantum,chakraborty2019power,kerenidis2020prakash}.
We assume a dense $N\times N$ matrix $\matA\in\mathbb{C}$, whose elements are known (i.e. pre-computed), and stored as an array in a classical memory. Given this classical array, a data structure of the form of \cite{kerenidis2016quantum} can be constructed in an \textit{offline} fashion and stored in a classical memory in $O(N^2)$ arithmetic operations, and it requires $O(N^2\log(N))$ space, i.e. words, assuming each matrix element requires one word to store.

For dense matrices, the data structure consists of $N$ binary trees with $O(\log(N))$ depth and $N$ leaves each. If we were to prepare a quantum circuit for the access oracle of this data structure, we would need $O(N^2)$ gates to implement it. To elaborate more, each binary tree $B_i$ can be obtained by constructing a circuit that rotates the state $\ket{0}^{\otimes \log(N)}$ to the state $\ket{B_i}$, where the latter contains the $N$ leaves of the tree $B_i$. Such a circuit can be constructed using $O(N)$ one and two qubit gates (rotations and CNOT gates) \cite{grover2002creating,mottonen2005transformation}, and we need such a circuit for every $i\in[N]$. However, if the quantum computer has the ability to query the elements of the classical data structure in superposition, which is possible in the QRAM model, the gate complexity can be reduced to $O(\polylog(N))$ \cite{kerenidis2016quantum}. 
In particular, Theorem 10 of~\cite{kerenidis2016quantum} showed that this construction enables efficient circuits implementing the maps
\begin{equation}
    \ket{i}\ket{0}\rightarrow\ket{i}\ket{\matA_i}\quad\text{and}\quad\ket{0}\ket{j}\rightarrow\ket{\matAtilde}\ket{j},
\end{equation} 
where $\ket{\matA_i}$ is the normalized quantum state corresponding to the $i$-th row of $\matA$ and $\braket{i|\matAtilde}=\norm{\matA_i}$ encodes the row norms.
These circuits can be implemented to precision $\epsilon$ with gate complexity $O\left(\polylog(N/\epsilon)\right)$~\cite{kerenidis2016quantum, chakraborty2019power}.
The following lemma can be used to construct a block-encoding from a quantum-accessible data structure in QRAM.

\begin{lemma}[Imported result from \cite{chakraborty2019power,gilyen2019quantum}]\label{lemma:frobenius_norm_block_encoding}
   Let $\matA\in\mathbb{C}^{N\times N}$ be a matrix stored in a quantum-accessible data structure. 
   There exists a $\QENC_{a,b}(\matA,\epsilon)$ subroutine with $a=\|\matA\|_F,b=\lceil \log(N)\rceil+1,$ and cost $\qcost(\matU_{\matA};\epsilon)=O(\polylog(N/\epsilon))$.
\end{lemma}

By using the specific block-encodings from Appendix \ref{appendix:preliminaries} and state preparation circuits from Section \ref{section:main_subroutines}, we can bound the complexity of the algorithm as follows.
\begin{corollary}
    \label{corollary:qgap_specific}
    Using \Cref{lemma:frobenius_norm_block_encoding} to construct block-encodings, and \Cref{lemma:oblivious_state_preparation} for state preparation, the total complexity of Theorem \ref{theorem:alg_qeig} becomes:
    \begin{align*}
        O\left(
        \tfrac{N^2 b}{\epsilon^{2}\gap_k^{2}}\polylog(N,\delta^{-1},\epsilon^{-1},\gap_k^{-1})
        \right),
    \end{align*}
    and the algorithm succeeds with probability at least $1-\delta$.
    \begin{proof}
        It suffices to replace the costs of the block-encodings in Theorem \ref{theorem:alg_qeig}. 
        For all block-encodings we can use \Cref{lemma:frobenius_norm_block_encoding}, which means that $a=\|\matH\|_F\in O(\sqrt{N})$. The cost has a polylogarithmic dependence on the accuracy. 
        Replacing the costs in Eq. \eqref{eq:main_theorem_cosst_with_specific_osp} we have
        \begin{align*}
            \qcost(\matU_{\matH};\epsilon_1)
            &\in \polylog(N/(\epsilon\gap_k)),
            \\
            \qcost(\matU_{\matH};\epsilon_2)
            &\in \polylog(N/\gap_k).
        \end{align*}
        Combining everything together, Eq. \eqref{eq:main_theorem_cosst_with_specific_osp} becomes
        \begin{align*}
        O\left(
            N^2 b
            \left(
                \frac{1}{\epsilon^2\gap_k^2}
                +
                \frac{1}{\gap_k^{2}} 
            \right)
            \polylog\left(N,a,\epsilon^{-1},\gap_k^{-1},\delta^{-1}\right)
        \right),
        \end{align*}
        where we absorbed all polylogarithmic factors in the trailing term.
    \end{proof}
\end{corollary}

\section{A lower bound in the Black-Box model}
\label{section:lower_bounds}
In this section we derive a lower bound for the problem of deciding the existence of a spectral gap of a binary matrix, in the so-called Black-Box access model. Here we assume a unitary $\matO_{\matA}$ through which we can query elements of the matrix $\matA$ via:
\begin{align*}
    \matO_{\matA}:\ket{i}\ket{j}\ket{0} \mapsto \ket{i}\ket{j}\ket{\matA_{i,j}}.
\end{align*}
Several lower bounds are known for related problems. 
A straightforward query lower bound for determining the midpoint $\mu_k$ can be obtained from the median problem \cite{nayak1999quantum}. Let $\mathcal{A}$ be a quantum algorithm which, given $\matO_{\matA}$, and an integer $k$, it returns any (classical) value $\widetilde\mu_k$ such that $\widetilde\mu_k\in(\lambda_k,\lambda_{k+1})$. Then $\mathcal{A}$ can be directly used to solve the median problem, where, given oracle access to a sequence of $N$ numbers $\{x_1,\ldots,x_N\}$, where $N$ is even, we seek to find a number that lies strictly inside $x\in(x_{N/2},x_{N/2}+1)$. Such a sequence can be seen as the elements of a diagonal matrix, and the eigenvalues of such a matrix are the diagonal elements themselves.
\cite{nayak1999quantum} proved a $\Omega(N)$ query lower bound for the median problem, assuming that we have oracle access to a quantum circuit implementing the sequence. Indeed, the same bound holds even if we only want to verify that $\widetilde\mu_k$ is inside the correct interval, since this generalizes the counting problem \cite{nayak1999quantum,aaronson2020quantum,aaronson2020quantumlower}.

Here we obtain a stronger lower bound for deciding whether there exists a gap between two consecutive eigenvalues, inspired by the reductions of \cite{dorn2009quantum,gribling2024optimal}.
We remark that the construction below is inherently non-Hermitian. Whether analogous black-box lower bounds hold for Hermitian matrices remains an open question.
\begin{theorem}
    Given a matrix $\matA\in\{0,1\}^{N\times N}$, a unitary $\matO_{\matA}:\ket{i}\ket{j}\ket{0} \mapsto \ket{i}\ket{j}\ket{\matA_{i,j}}$, and an integer $k$, any quantum algorithm which answers TRUE if $\gap_k:=|\lambda_k(\matA)-\lambda_{k+1}(\matA)|> 0$, and FALSE otherwise, must execute $\Omega(N^2)$ queries to $\matO_{\matA}$. Here the eigenvalues of $\matA$ are order such that $
            Re(\lambda_1)\geq Re(\lambda_2) \geq \ldots \geq Re(\lambda_N) 
    $, and if $Re(\lambda_k)=Re(\lambda_{k+1})$, then $Im(\lambda_k)\geq Im(\lambda_{k+1}).$
    \begin{proof}
        The proof closely follows the techniques of \cite{dorn2009quantum,gribling2024optimal} for other similar problems (rank, determinant, matrix power, and inverse), essentially by reducing the problem to counting the number of triangles in a directed graph. Assume that we are given a set $X\in\{0,1\}^{N^2}$ and the value $a=\lceil N^2/2 \rceil$, and we ask whether $\sum_{i=1}^{N^2} X_i=a$. The query complexity of this problem is $\Omega(N^2)$ \cite{dorn2009quantum}.
        
        For the reduction the first step is to create a couple of graphs. We create a bipartite graph $G$, with $N$ sources $s_i$ and $N$ targets $t_j$, and we insert an edge between $s_i$ and $t_j$ iff $X_{iN+j}=1$. We also create a utility node $v$, which has $N$ outgoing edges to every one of the sources, and $N$ incoming edges from every one of the targets.

        Thereafter, we construct a second bipartite graph $G'$ which also has $N$ sources $s'_i$, $N$ targets $t'_j$, and a utility node $v'$. We create exactly $a$ edges between the sources and the targets, in a round-robin manner, so that the edges are ``evenly distributed'' among the two vertex sets (for each $s_i$, we add roughly $\lceil a/N \rceil$ outgoing edges). Finally, we add $N$ edges from $v'$ to every $s'_i$, and $N$ edges from every $t'_j$ to $v'$.

        Note that the two graphs are disjoint. Consider the larger graph $F=(V\cup V',E\cup E')$, and its adjacency matrix $\matA$. $F$ has $4N+2$ nodes. If we choose the node labels of $F$ appropriately, $\matA$ consists of two diagonal blocks: one block which corresponds to the adjacency of $G$, and the other to the one of $G'$. More specifically,
        \begin{align*}
            \matA = \begin{pmatrix}
                \matA_G & \\
                 & \matA_{G'}
            \end{pmatrix}.
        \end{align*}
        We will show that $\matA_G$ and $\matA_{G'}$ have only three nonzero eigenvalues.
        
        The elements of the matrix $\matA^3$ contain the number of paths of length three starting from every node $i$ to every other node $j$. 
        Since the only length-three paths from $v$ and $v'$ are towards themselves, the matrix $\matA^3$ has the following structure:
        \begin{align*}
            \matA^3
            =
            \begin{pmatrix}
                \sigma &  & & \\
                 & \matB & & \\
                 & & a & \\
                 & & & \matB'
            \end{pmatrix},
        \end{align*}
        where $a$ is located in position $\matA^3_{v',v'}$ and $\sigma:=\sum_{i=1}^{N^2}X_i$ in position $\matA^3_{v,v}$. It is not hard to see that $\sigma$ and $a$ are eigenvalues, since $\det(\sigma\matI-\matA^3)=\det(a\matI-\matA^3)=0$.

        The matrices $\matB$ and $\matB'$ have a special structure. Let us analyze $\matB$ first. From each source vertex $s_i,i\in[N]$, there are exactly $d_i=outdeg(s_i)$ paths of length three to every  source $s_{j},$ including itself. This means that the row $i$ of $\matB$ contains exactly $N$ non-zeros, with the same value $d_i$, at the positions corresponding to the sources. This gives rise to a submatrix $\matB_s$. Looking at the targets $t_l$, note that there are exactly $f_l=indeg(t_l)$ paths from every node $t_m,m\in[N]$, to $t_l$. This means that the $l$-th column of the matrix $\matB$ has exactly $N$ nonzeros, equal to $f_l$, at the positions that correspond to target vertices. This gives rise to a submatrix $\matB_t$, where $\matB_s$ and $\matB_t$ are disjoint. With this description, we conclude that the matrix $\matB$ has the following structure:
        \begin{align*}
            \matB = \begin{pmatrix}
                \matB_s & \\
                & \matB_t
            \end{pmatrix},
        \end{align*}
        where both $\matB_s$ and $\matB_t$ have size $N\times N$. It turns out that they are rank-one matrices. Indeed, we can write $\matB_s=\vecd_s\vece^\top$, where $\vece$ is the all-ones vector, $\vecd_s^\top =\begin{pmatrix} d_1 & \ldots & d_N \end{pmatrix}$, and $\matB_t=\vece\vecf_t^\top$, where $\vecf_t^\top =\begin{pmatrix} f_1 & \ldots & f_N \end{pmatrix}$. The only nonzero eigenvalues of $\matB_s$ and $\matB_t$ are equal to $\lambda_s=\vecd_s^\top \vece=\sum_{i=1}^N d_i$ and $\lambda_f=\vecf_t^\top \vece=\sum_{i=1}^N{f_i}$. We can see that
        \begin{align*}
            \sigma = \lambda_s = \lambda_f.
        \end{align*}

        Next, we shift our attention to $\matB'$. Since it has very similar structure to $\matB$, we can argue accordingly that the only two non-zero eigenvalues of $\matB'$ are equal to $a$. Now, it follows that \begin{align*}
            \Lambda(\matA^3)
            =
            \{a\}\cup\{\sigma\}\cup\Lambda(\matB)\cup\Lambda(\matB')
            =\{a,a,a,\sigma,\sigma,\sigma\}.
        \end{align*}
        The above imply that $\matA^3$ is diagonalizable.
        The eigenvalues of $\matA$ are the third-roots of the eigenvalues of $\matA^3$. Since $\matA$ is real, its complex eigenvalues come in conjugate pairs. The spectrum of $\matA$ is equal to
        \begin{align*}
            \Lambda(\matA) = \{a_1,a_2,a_3,\sigma_1,\sigma_2,\sigma_3\},
        \end{align*}
        where $\{a_1,a_2,a_3\}$ are either all equal to $|a^{1/3}|$ or equal to $|a^{1/3}|\times \{z_1,z_2,z_3\}$ where $z_1,z_2,z_3$ are the three third-roots or unity. Similarly,  $\{\sigma_1,\sigma_2,\sigma_3\}$ are either all equal to $|\sigma^{1/3}|$ or $|\sigma^{1/3}|\times \{z_1,z_2,z_3\}$.

        Let $\{\lambda_1,\lambda_2,\ldots,\lambda_6\}$ be the eigenvalues of $\matA$ sorted in descending order of their real parts, where conflicts are resolved with respect to the imaginary part, i.e.
        \begin{align*}
            Re(\lambda_1)\geq Re(\lambda_2) \geq \ldots \geq Re(\lambda_6).
        \end{align*}
        We now derive the final argument.\\
        $(\implies):$ If $a= \sigma$, we have the following four cases for the eigenvalues of $\matA$:
        \begin{center}
        \begin{tabular}{c  c  c}
            $a$ & $\sigma$ &  Relation\\\hline
            Real & Real &  $\lambda_i=\lambda_j$, $j\in[6]$.
            \\
            Real & Complex &  $\lambda_1=\lambda_2=\lambda_3=\lambda_4\neq \lambda_5\neq \lambda_6$. 
            \\
            Complex & Real &  Same as above. \\
            Complex & Complex &  $\lambda_1=\lambda_2\neq \lambda_3=\lambda_4\neq\lambda_5=\lambda_6$.
        \end{tabular}    
        \end{center}
        $(\impliedby):$ On the opposite direction, we prove each case separately:
        \begin{itemize}
            \item If $\lambda_i=\lambda_{i+1}$, for $i=1,\ldots,5$, all eigenvalues are equal and hence $a=\sigma$.
            \item $\lambda_1=\lambda_2=\lambda_3=\lambda_4\neq \lambda_5\neq \lambda_6$. We already argued that $a_i$ are either all equal to each other or equal to the scaled roots of unity. The same for $\sigma_i$. The only way that the top-three $\lambda_i$ are equal is if either all $a_i$ are equal and real, or all $\sigma_i$. Since they are real, and equal to $\lambda_4$, it means that $\lambda_4$ is also real. We conclude that at least one of $\lambda_1,\ldots,\lambda_4$ is equal to $a$, and at least one of them is equal to $\sigma$, therefore $a=\sigma$.
            \item The only way that top-6 eigenvalues are equal in pairs, is if both $a_i$ and $\sigma_i$ are the scaled roots of unity, and, in this case, with the same scaling factor, which also implies that $a=\sigma$.
        \end{itemize}
        We conclude that $a=\sigma$ if and only if one of the following relation holds:
        \begin{itemize}
            \item $\lambda_i=\lambda_{i+1},i=1,\ldots,5.$
            \item $\lambda_1=\lambda_2=\lambda_3=\lambda_4\neq \lambda_5\neq \lambda_6$.
            \item $\lambda_1=\lambda_2\neq \lambda_3=\lambda_4\neq\lambda_5=\lambda_6$.
        \end{itemize}
        Therefore, one can call the gap algorithm five times, for $k=1,\ldots,5$, and revisit the three cases above in order to verify whether $a=\sigma$.
    \end{proof}
\end{theorem}

\section{Conclusion}
\label{section:conclusion}
In this work we studied algorithms for computing spectral gaps between eigenvalues of Hermitian matrices and the corresponding midpoints. One of the main subroutines is a randomized state preparation method, which achieves at least $1/N$ overlap with any fixed target state with probability at least $3/5$, and requires $O(N)$ classical random bits to prepare. This can be used to efficiently approximate the smallest singular value of a block-encoded matrix with high probability, and to count the eigenvalues of a Hermitian matrix that are smaller than a given threshold. In the quantum-accessible data structure model, the total query complexity of our main algorithm scales at most quadratically with respect to the problem size $N$. This indicates the potential for a polynomial quantum speed-up, since the complexity of the best-known classical algorithms require $\widetilde O(N^{\omega})$ arithmetic operations. In the black-box access model, we also proved an $\Omega(N^2)$ query lower bound for deciding whether there is a gap between two consecutive eigenvalues of a binary matrix.

\bibliographystyle{quantum}

\appendix

\section{Preliminaries on block-encodings and eigenvalue transformation}
\label{appendix:preliminaries}
In this section we recall some basic definitions and known results for block-encodings.
Block-encoding refers to the derivation of a unitary operator $\matU_{\matA}$ which can be described by a quantum circuit and ``encodes'' a given matrix $\matA$, which was stated in Definition \ref{definition:block_encoding}.

Constructing block-encodings with low cost is not a trivial task. A common approach in the literature is to assume a ``black-box'' unitary oracle that gives access to a (sparse) matrix $\matA$. See eg. \cite{gilyen2019quantum,martyn2021grand}. Recent works take a step further, describing explicit quantum circuits for the block-encoding of special classes of matrices \cite{camps2024explicit,sunderhauf2024block,chakraborty2019power,camps2022fable}. In the main algorithm we assume the existence of a general subroutine that can prepare block-encodings, defined as follows. 
\begin{definition}
    \label{definition:block_encoding_subroutine}
    A subroutine $\matU_{\matA} \gets \QENC_{a,b}(\matA,\epsilon)$ takes as input a matrix $\matA$ of size $N$ and an accuracy $\epsilon$, and prepares an $(a,b,\epsilon)$-QUBE of $\matA$. $a$ and $b$ might be functions of $N$ or $\|\matA\|_\xi$, for some norm $\xi$.
\end{definition}
It is straightforward to obtain a block-encoding after a diagonal shift.
\begin{lemma}[$\QSHIFT$, follows from \cite{gilyen2019quantum}]
    \label{lemma:sgn_new}
    Let $h\in[-1/2,1/2]$ be a scalar and $\matH$ be a Hermitian matrix with $\|\matH\|\leq 1/2$, and  $\matU_{\matH}$ be an $(a,b,\epsilon_{\rm enc})$-QUBE for $\matH$ with cost $\qcost(\matU_{\matH};\epsilon_{\rm enc})$. We can construct a $(a+|h|,b+1,\epsilon_{\rm enc})$-QUBE for $\matH-h$ with a subroutine 
    \begin{align*}
        \matU_{\matH-h}\gets \QSHIFT(\matU_{\matH},h,a),
    \end{align*} where $\qcost(\matU_{\matH-h};\epsilon_{\rm enc})=\qcost(\matU_{\matH};\epsilon_{\rm enc})+O(1)$.
    \begin{proof}
        We use \cite[Lemma 29]{gilyen2019quantum} in a similar way as in \cite{Lin2020nearoptimalground}. First we construct a couple of $2$-by-$2$ unitaries $\matP_L\ket{0}=c_1\ket{0}+c_2\ket{1}$ and $\matP_R\ket{0}=d_1\ket{0}+d_2\ket{1}$, as follows. Set $\beta=1+|h|/a$. The values $c_1,d_1,c_2,d_2$ are chosen such that $\beta(c_1^*d_1)=1$ and  $\beta(c_2^*d_2)=-h/a$. Let $\matW=\ket{0}\bra{0}\otimes \matU_{\matH} + 
        \ket{1}\bra{1}\otimes \matU_{\matH}$. One can verify that $(\matP_L^\dagger\otimes \matI) \matW (\matP_R\otimes \matI)$ is an $(a+|h|,b+1,\epsilon_{\rm enc})$-QUBE of $\matH-h$.
    \end{proof}
    
\end{lemma}

\subsection{Quantum eigenvalue transformation and matrix sign function}
We can now describe how to prepare a block-encoding of the matrix sign function $\sgn(\matH-h)$ using  Quantum Eigenvalue Transformation (QET). We follow the methodology of \cite{Lin2020nearoptimalground} which is convenient for our analysis, but we also note that many works have studied similar approximations \cite{martyn2021grand,gilyen2019quantum,mizuta2024recursive,low2017quantum}. E.g., an elegant an practical approach is mentioned \cite{mizuta2024recursive}, using a recursive  QET based on Pad\'e approximations for the sign function, with sharp complexity bounds.

\begin{corollary}[Adapted from \cite{Lin2020nearoptimalground,gilyen2019quantum}]
    \label{corollary:sign_block_encoding}
    Let $\matU_{\matH}$ be an $(a,b,\epsilon_{\rm enc})$-QUBE of a Hermitian matrix $\matH$, $\|\matH\|\leq 1/2$, with gate complexity $\qcost(\matU_{\matH};\epsilon_{\rm enc})$. Let $h$ be a scalar in $[-1/2,1/2]$, and assume that $\Lambda\left(
        \matH-h
    \right)\subseteq [-1,-\Delta]\cup[\Delta,1]$, for some (known) $\Delta\in(0,1)$. Let $\epsilon_{\rm sgn}\in(0,1)$ be an input accuracy parameter. 
    If $\epsilon_{\rm enc}\leq \epsilon_{\rm sgn}\frac{\Delta^2}{C_{\rm sgn}(a+|h|)^2}$, for some absolute constant $C_{\rm sgn}>1$, then we can construct a unitary $\matU_{\rm sgn}$ which is an $(1,b+2,\epsilon_{\rm sgn})$-QUBE of $\sgn(\matH-h)$. 
    The cost of $\matU_{\rm sgn}$ is:
    \begin{enumerate}
        \item Queries to $\matU_{\matH -h}$: $O\left(
            \frac{a+|h|}{\Delta}\log(
                \frac{1}{\epsilon_{\rm sgn}}
            )
        \right)$, with $\qcost(\matU_{\matH-h};\epsilon_{\rm enc})=\qcost(\matU_{\matH};\epsilon_{\rm enc})$,
        \item Number of qubits: $\log(N)+b+2$,
        \item Other elementary gates: $O\left(
            \frac{a+|h|}{\Delta}b\log(
                \frac{1}{\epsilon_{\rm sgn}}
            )
        \right)$
        \item Total complexity: $C({\matU_{\rm sgn};\Delta,\epsilon)} = O\left( (\qcost(\matU_{\matH};\epsilon_{\rm enc})+b)\cdot\frac{a+|h|}{\Delta}\log(\frac{1}{\epsilon_{\rm sgn}})\right)$.
    \end{enumerate}
    \begin{proof}
        Since $\matU_{\matH}$ is an $(a,b,\epsilon_{\rm enc})$-QUBE for $\matH$, then, by definition, it is also an $(1,b,0)$-QUBE for a matrix $\matHtilde$, where $\|\matH-a\matHtilde\|\leq\epsilon_{\rm enc}$. We use Lemma \ref{lemma:sgn_new} to prepare an $(a+|h|,b+1,\epsilon_{\rm enc})$-QUBE $\matU_{\matH-h}$ for $\matH-h$ with the same cost (up to constants). Again, by definition, $\matU_{\matH-h}$ is an $(1,b+1,0)$-QUBE for a matrix $\matHtilde'$ such that $\|(\matH-h)-(a+|h|)\matHtilde'\|\leq \epsilon_{\rm enc}$. 
        
        Using standard techniques from holomorphic functional calculus (see e.g. \cite[Lemma B.1]{sobczyk2024invariant}) we know that as long as $\epsilon_{\rm enc}\leq \frac{\epsilon_{\rm sgn}}{2}\cdot \frac{\sigma_{\min}^2(\matH-h)}{C_{\rm sgn}(a+|h|)}$, for some constant $C$, then it holds that $\left\|\sgn\left(\frac{\matH-h}{a+|h|}\right)-\sgn(\matHtilde')\right\|\leq \epsilon_{\rm sgn}/2$. 
        A sufficient upper bound for $\epsilon_{\rm enc}$ is to set $\epsilon_{\rm enc}\leq\epsilon_{\rm sgn}\frac{\Delta^2}{2C_{\rm sgn}(a+|h|)}$. It also holds that $\sgn\left(\frac{\matH-h}{a+|h|}\right)=\sgn\left(\matH-h\right)$ since $\sgn$ is scale invariant. 
        
        It remains to approximate $\sgn(\matHtilde')$ with a polynomial (QET).
        From Weyl's inequality, $|\lambda_i(\matH-h)-(a+|h|)\lambda_i(\matHtilde')|\leq \epsilon_{\rm enc}$. Therefore, it follows that $\sigma_{\min}(\matHtilde')\geq \frac{ \sigma_{\min}(\matH-h)-\epsilon_{\rm enc}}{a+|h|}
        \geq
        \frac{ \Delta-\epsilon_{\rm enc}}{a+|h|}
        \gg
        \frac{ \Delta}{2(a+|h|)}
        $, where the last bound holds due to the aforementioned value of $\epsilon_{\rm enc}\leq\epsilon_{\rm sgn}\frac{\Delta^2}{2C(a+|h|)}\ll \Delta/2$. 
        From \cite[Lemma 3]{Lin2020nearoptimalground}, we can find a polynomial $p(\cdot)$ with degree $l=O\left(
            \frac{2(a+|h|)}{\Delta}\log(
                \frac{2}{\epsilon_{\rm sgn}}
            )
        \right)$, such that $|p(x)|\leq 1$ for all $|x|\leq 1$, and $|p(x)-\sgn(x)|\leq \epsilon_{\rm sgn}/2$ for all $x\in[-1,-\Delta/2(a+|h|)]\cup [\Delta/2(a+|h|),1]$. 
        Consequently, $\left\|p(\matHtilde')-\sgn(\matHtilde')\right\|\leq \epsilon_{\rm sgn}/2$. 
        
        From \cite[Thm. 2]{gilyen2019quantum}  (or \cite[Thm. 1]{Lin2020nearoptimalground}), we can obtain an $(1,b+2,0)$-QUBE of $p(\matHtilde')$ using $l$ queries of $\matU_{\matH-h},\matU_{\matH-h}^\dagger$, and $O((b+2)l)$ elementary gates. This is $\matU_{\rm sgn}$, and it is easy to see that it forms an $(1,b+2,\epsilon_{\rm sgn})$-QUBE for $\sgn(\matH-h)$, since:
        \begingroup
        \allowdisplaybreaks
        \begin{align*}
            \left\|
                p(\matHtilde')-\sgn(\matH-h)
            \right\|
            &=
            \left\|
                p(\matHtilde')-\sgn(\tfrac{\matH-h}{a+|h|})
            \right\|
            \\
            &\leq
            \left\|
                p(\matHtilde')
                -
                \sgn(\matHtilde')
            \right\|
            +
            \left\|
            \sgn(\matHtilde')
            -\sgn(\tfrac{\matH-h}{a+|h|})
            \right\|
            \\
            &\leq
            \epsilon_{\rm sgn}/2 + \epsilon_{\rm sgn}/2
            \\
            &= \epsilon_{\rm sgn}.
        \end{align*}\endgroup\end{proof}
\end{corollary}

\subsection{Computing the trace of a block-encoded matrix}
The main algorithm requires a subroutine to compute the trace of a block-encoded matrix.

\begin{definition}[\cite{rall2020}]
    Let $\rho$ be a density operator on $\mathcal{H}_S$ and let $\ket{\mathbf{0}}$ be some easy-to-prepare state in $\mathcal{H}_S$.
    A unitary $\matU_\rho$ on $\mathcal{H}_S\otimes\mathcal{H}_A$ is an $R$-preparation unitary of $\rho$ if we have
    \begin{equation}
        \rho = \tr_{\mathcal{H}_A}\left(\ket{\rho}\bra{\rho}\right)
    \end{equation}
    where $\ket{\rho}=\matU_\rho\left(\ket{\mathbf{0}}\ket{\mathbf{0}}\right)$ and $\matU_\rho$ is implementable using $R$ elementary gates.
\end{definition}

\begin{proposition}\label{prop:id_purification}
    Let $\matI_n$ denote the $2^n \times 2^n$ identity matrix.
    Then there exists a $2n$-preparation of $\frac{1}{2^n}I_n$ using $n$ CNOT gates and $n$ single-qubit gates on $\mathcal{H}\otimes \mathcal{H}$.
\end{proposition}
\begin{proof}
    Let $\ket{i}_n$ denote the $i^{\text{th}}$ basis state on the $2^n$-dimensional Hilbert space.
    Then we can prepare the state
    \begin{equation}
        \ket{\rho}=\frac{1}{\sqrt{2^n}}\sum_{i=0}^{2^n -1} \ket{i}_n \ket{i}_n
    \end{equation}
    by first applying $\matH^{\otimes n}\otimes I_n$, where $\matH$ denotes the Hadamard gate, and then CNOT gates between the $j^{\text{th}}$ and the $(j+n)^{\text{th}}$ qubits for $j=0,\cdots , n-1$.
    Then, we have
    \begin{align}
        \tr_{\mathcal{H}}\left(\ket{\rho}\bra{\rho}\right) &= \tr_{\mathcal{H}}\left(\frac{1}{2^n}\sum_{i,j=0}^{2^n -1} \ket{i}_n \ket{i}_n\bra{j}_n \bra{j}_n\right)
        = \frac{1}{2^n}\sum_{i,j=0}^{2^n -1}\ket{i}_n\bra{j}_n\tr_{\mathcal{H}}\left(\ket{i}_n\bra{j}_n \right)\\
        &= \frac{1}{2^n}\sum_{i=0}^{2^n -1}\ket{i}_n\bra{i}_n = \frac{1}{2^n} I_n.
    \end{align}
    
\end{proof}

\begin{corollary}\label{corollary:quantum_trace_p}
    Let $\matA$ be an $N \times N$ matrix, and $\delta_{\rm tr},\epsilon_{\rm tr}\in(0,1)$. Let  $\matU_{\matA}$ be an $(a,b,\tfrac{\epsilon_{\rm tr}}{2N})$-QUBE of $\matA$ with circuit complexity $\qcost(\matU_{\matA};\frac{\epsilon_{\rm tr}}{2N})$ such that
    \begin{align*}
        \lVert \matA - a\matAtilde\rVert\leq \tfrac{\epsilon_{\rm tr}}{2N},
    \end{align*}
    where $\matAtilde:=(\bra{0}^{\otimes b} \otimes \matI)\matU_{\matA}(\ket{0}^{\otimes b} \otimes \matI)$. There exists a quantum algorithm that produces an estimate $\xi$ of $\tr(\matA)$ such that
    \begin{align*}
        \lvert \xi - \tr(\matA) \rvert \leq \epsilon_{\rm tr},
    \end{align*}
    with probability at least $(1-\delta_{\rm tr})$.
    The algorithm has complexity:
    \begin{enumerate}
        \item Queries to $\matU_{\matA}$: $O\left(
            \tfrac{aN}{\epsilon_{\rm tr}}
        \log(\tfrac{1}{\delta_{\rm tr}})\right)$,
        \item Number of qubits: $2\log(N)+b$,
        \item Other elementary gates: $O\left(
            \log(N)\tfrac{aN}{\epsilon_{\rm tr}}
        \log(\tfrac{1}{\delta_{\rm tr}})\right)$
        \item Total complexity: $O\left((\log(N)+\qcost(\matU_{\matA};\tfrac{\epsilon_{\rm tr}}{2N}))
        \tfrac{aN}{\epsilon_{\rm tr}}
        \log(\tfrac{1}{\delta_{\rm tr}})\right)$. 
    \end{enumerate}
    \begin{proof}
    From Proposition~\ref{prop:id_purification}, there exists an $R$-preparation of $\rho = \frac{1}{N}I_n$ with cost $R=O(\log(N))$ elementary gates. From Lemma~5 of~\cite{rall2020}, we know that if  $\matU_{\matA}$ is an $(a,b,0)$-QUBE of $\matA$ with circuit complexity $\qcost(\matU_{\matA})$, and $\rho$ has an $R$-preparation unitary, then for every $\epsilon,\delta>0$ there exists an algorithm that produces an estimate $\xi$ of $\tr(\rho \matA)$ such that
    \begin{equation}
        \lvert \xi - \tr(\rho \matA) \rvert \leq \epsilon
    \end{equation}
    with probability at least $(1-\delta)$.
    The algorithm uses $O\left(\frac{a}{\epsilon}\log(\frac{1}{\delta})\right)$ applications of a Grover operator that requires four uses of the R-preparation unitary and two queries to $\matU_{\matA}$.
    Note that $\matU_{\matA}$ is an $(1,b,0)$-QUBE of $\matAtilde$. Using \cite[Lemma 5]{rall2020} on $\matAtilde$ with error $\epsilon:=\tfrac{\epsilon_{\rm tr}}{2aN}$, we can compute an estimate $\xi'$ such that
    \begin{align*}
        \labs \xi' - \tr(\tfrac{1}{N}\matAtilde) \rabs \leq \epsilon,
    \end{align*}
    with probability at least $1-\delta_{\rm tr}$ and circuit complexity 
    \begin{align*}
        O\left((\log(N)+\qcost(\matU_{\matA}))\frac{1}{\epsilon}\log(\frac{1}{\delta_{\rm tr}})\right) 
        &= 
        O\left((\log(N)+\qcost(\matU_{\matA}))\frac{aN}{\epsilon_{\rm tr}}\log(\frac{1}{\delta_{\rm tr}})\right).
    \end{align*}
    Then we can set $\xi=a N\xi'$, which satisfies
    \begingroup
    \allowdisplaybreaks
    \begin{align*}
        \labs \xi - \tr(\matA) \rabs 
        &\leq 
        \labs \xi - \tr(a \matAtilde) \rabs 
        + 
        \labs \tr(a\matAtilde) - \tr(\matA) \rabs \\
        &= 
        aN\labs \xi' - \tr(\tfrac{1}{N}\matAtilde) \rabs 
        + 
        \labs \tr(a\matAtilde) - \tr(\matA) \rabs \\
        &\leq 
        aN\tfrac{\epsilon_{\rm tr}}{2aN} 
        + 
        N\lnorm a\matAtilde - \matA \rnorm
        \\
        &\leq \epsilon_{\rm tr}.
    \end{align*}
    \endgroup
\end{proof}
\end{corollary}

\end{document}